\documentclass[fleqn,10pt]{wlscirep}
\usepackage[utf8]{inputenc}
\usepackage[T1]{fontenc}
\usepackage{amsmath}
\usepackage{textcomp}
\usepackage{gensymb}
\usepackage{float}
\usepackage{lipsum}
\usepackage{graphicx}
\usepackage{bm}

\title{Magneto-electric decoupling in bismuth ferrite}

\author[1,*]{Thien Thanh Dang}
\author[1,2]{Juliana Heiniger-Schell}
\author[1]{Astita Dubey}
\author[3]{João Nuno Gonçalves}
\author[1]{Marianela Escobar Castillo}
\author[1]{Daniil Lewin}
\author[1]{Ian Chang Jie Yap}
\author[4]{Adeleh Mokhles Gerami}
\author[1]{Sobhan Mohammadi Fathabad}
\author[5]{Dmitry Zyabkin}
\author[1]{Doru Constantin Lupascu}

\affil[1]{Institute for Materials Science and Center for Nanointegration Duisburg-Essen (CENIDE), University of Duisburg-Essen, 45141 Essen, Germany}
\affil[2]{European Organization for Nuclear Research (CERN), CH-1211 Geneva, Switzerland}
\affil[3]{CICECO—Aveiro Institute of Materials and Departamento de Física, Universidade de Aveiro, 3810-193 Aveiro, Portugal}
\affil[4]{School of Particles and Accelerators, Institute for Research in Fundamental Sciences (IPM), P.O. Box 19395-5531, Tehran, Iran}
\affil[5]{Chair Materials for Electronics, Institute of Materials Science and Engineering, and Institute of Micro and Nanotechnologies MacroNano®, TU Ilmenau, 98693 Ilmenau, Germany}
\affil[*]{e-mail: thien.dang@uni-due.de}

\begin{abstract}
It is still under intensive discussion how magnetoelectric coupling actually occurs at the atomic scale in multiferroic BiFeO$_3$. Nuclear solid-state techniques monitor local fields at the atomic scale. Using such an approach, we show that, contrary to our own expectation, ferroelectric and magnetic ordering in bismuth ferrite (BiFeO$_3$ or BFO) decouple at the unit-cell level. Time differential perturbed angular correlation (TDPAC) data at temperatures below, close, and above the magnetic Néel temperature show that the coupling of the ferroelectric order to magnetization is completely absent at the bismuth site. It is common understanding that the antiferromagnetic order and the cycloidal ordering due to the Dzyaloshinskii-Moriya interaction generate a net zero magnetization of the sample, cancelling any magnetoelectric effect at the macroscopic level. Our previous data show that a very large coupling of magnetic moment and electrical distortions arises on the magnetic sub-lattice (Fe-site). The oxygen octahedra around the iron site experience a large tilt due to the onset of magnetic ordering. Nevertheless, the Bi-containing complementary sub-lattice carrying the ferroelectric order is practically unaffected by this large structural change in its direct vicinity. The magnetoelectric coupling thus vanishes already at the unit cell level. These experimental results agree well with an \textit{ab-initio} density functional theory (DFT) calculation.

\end{abstract}

\begin{document}

\flushbottom
\maketitle

\thispagestyle{empty}

\section*{Introduction}

Bulk bismuth ferrite (BiFeO$_3$ or simply BFO) is a type-I multiferroic \cite{Khomskii2009}, exhibiting antiferrodistortive (AFD) order at temperatures below 1200 K, ferroelectric (FE) order with a large spontaneous polarization below the Curie temperature $T_C \approx 1100$ K, and antiferromagnetic (AFM) order below its Néel temperature $T_N \approx 650$ K \cite{Karpinsky2017, Fischer1980, Catalan2009}. The antiferromagnetic Néel temperature is much lower than its ferroelectric Curie temperature, but both order parameters concurrently prevail at room temperature. Thus, significant magnetoelectric coupling is expected at room temperature, making BFO potentially useful for applications in sensors, magnetoelectric memories, spintronics, and photovoltaics \cite{Puhan2021}.

Much research has been conducted to investigate BFO at the macroscopic scale utilizing numerous methods including electric polarization measurements \cite{Karpinsky2017}, piezoelectric testing \cite{Stevenson2015}, magnetic measurements using SQUID magnetometry \cite{Park2007}, or at the microscopic scale piezoelectric force microscopy (PFM) and photoemission electron microscopy (PEEM) based on X-ray linear dichroism (XLD) \cite{Zhao2006}, among others. However, these macroscopic investigations are limited due to the antiferromagnetic ordering of the material which diminishes the measurable magnetic signal.

\begin{figure*}[!htb]
\centering
\includegraphics[width=13cm]{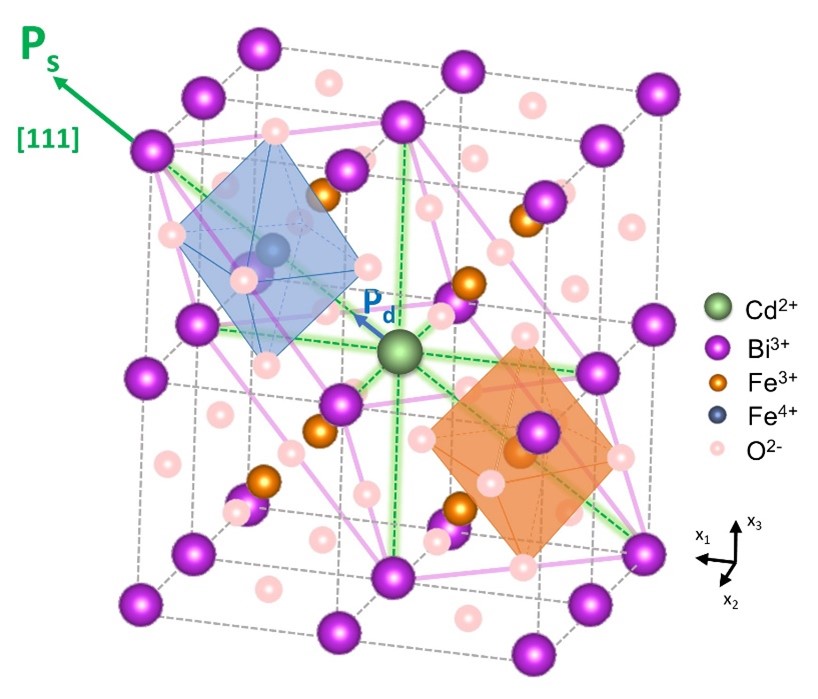}
\caption{The rhombohedral unit cell (formed by 2$\times$2$\times$2 supercell containing eight unit cells of the pseudo-cubic perovskite) of the crystal structure of BFO with Bi-site substitution of Cd$^{2+}$ and formation of Fe$^{4+}$. $P_s$ and $P_d$ denote the spontaneous ferroelectric polarization and the defect polarization generated by the defect dipole moment, respectively. The two oxygen octahedra (orange and blue) are always tilted and counter-rotated along the [111] direction in the AFD phase \cite{Ederer2005, Karpinsky2017}. These effects are not shown in this figure for simplicity.
}
\label{fig1}
\end{figure*}

A thorough comprehension of the structure of BFO at the atomic level is a challenging task due to its multiple ferroic ordering mechanisms within the same phase. The hybridization between Bi-6s and O-2p orbitals leads to a displacement of the Bi$^{3+}$ along the [111] pseudo cubic direction, producing the spontaneous ferroelectric polarization \cite{Kubel1990}. Antiferrodistortion (AFD) is defined as the static rotation of some atomic groups with respect to other parts of the crystal \cite{Gopalan2011}. In this work, the term AFD symmetry refers to the static rotation and tilting of FeO$_6$ octahedra along the [111] direction in opposite directions in adjacent unit cells \cite{Ederer2005, Karpinsky2017}. The antiferrodistortion generates lattice fluctuations, but no net macroscopic distortion \cite{Fechner2024}. With regard to the symmetry groups, the polar displacements alone would result in a reduction of the symmetry of the ideal perovskite structure (Pm$\bar{3}$m) to the rhombohedral space group R3m, whereas the rotation of the oxygen octahedra (a$^{-}$a$^{-}$a$^{-}$) alone would lead to the space group R$\bar{3}$c\cite{Benedek 2013}. The combination of both distortions results in the actual space group of BiFeO$_3$, R3c \cite{Ederer2005}. The rhombohedrally distorted perovskite structure of space group R3c results from 1 out of 8 possible orientations for polarization along the body diagonals of a pseudocubic unit cell. One such configuration is shown in Fig.\ref{fig1}. The anti-parallel iron spin alignment results in G-type antiferromagnetic (AF) ordering below the Néel temperature $T_N$ \cite{Kubel1990, Catalan2009}. The antiparallel order is slightly canted due to the Dzialoshinskii-Moriya \cite{Dzialoshinskii1957, Moriya1960} interaction yielding a small net magnetization per pair of spins. These net magnetizations in turn form an incommensurate cycloid annihilating all macroscopic magnetization.

On the unit cell level, NMR and Mössbauer studies both rely on $^{57}$Fe \cite{Zalessky2000, Landers2014, Gervits2022} where even the spin cycloid of BiFeO$_3$ can be recognized \cite{Landers2014}. Even though Gervits et al. have identified a signal from $^{209}$Bi, no further conclusions have been drawn from this signal. One technique that has proven to be useful in characterizing materials in general or specifically BFO at the atomic level is the time differential perturbed angular correlation (TDPAC) spectroscopy, which is another nuclear solid-state technique. The TDPAC method characterizes materials at the unit cell level by employing tracer probe ions \cite{Rasera1993}. These tracers do not typically affect the properties of the surrounding lattice. If chemically different from the host lattice, they represent a local defect that may attract electronic charge carriers, vacancies, or interstitial ions. Unlike Mössbauer spectroscopy, the signal in TDPAC is inherently temperature-independent, as no Debye-Waller factor is involved in spectrum formation \cite{Deicher1992}. If several TDPAC probes are available, certain ones can be intentionally positioned at various lattice sites due to e.g. their different ionic radii. For BFO, $^{111m}$Cd occupies the Bi site \cite{Marschick2020}, and $^{111}$In and $^{181}$Hf localize at the Fe site \cite{Schell2022, Dang2022}. As a result, the proper use of TDPAC probes can be instrumental in investigating BFO at both the Bi and Fe sites, whereas Mössbauer probes are limited to exploring the Fe site only. For example, the significant coupling between the electric and magnetic hyperfine fields observed in the magnetic atom Fe$^{3+}$ in the anti-ferromagnetic (AF) phase of BFO has been investigated by the time differential perturbed angular correlation (TDPAC) spectroscopy technique with $^{111}$In and $^{181}$Hf probes as tracer ions \cite{Schell2022, Dang2022}. The change of local probe environment at the phase transitions is well recognizable and determines the Néel and the Curie temperatures from within the unit cell to be approximately 370 °C and 829 °C, respectively. To investigate the local fields at the non-magnetic atom Bi$^{3+}$, $^{111m}$Cd was utilized. A first series of TDPAC measurements was conducted around the Curie point well above the Néel temperature (500 – 850 °C) \cite{Marschick2020}. That study determined the structural phase transition from rhombohedral $\alpha$-BFO in R3c setting to orthorhombic $\beta$-BFO with its Pbnm space group that occurs at 820 °C. \textit{Ab-initio} DFT simulations confirmed the substitution of Cd$^{2+}$ at the Bi$^{3+}$ site \cite{Marschick2020}. There is an open question regarding whether magnetoelectric coupling occurs to the non-magnetic atom Bi$^{3+}$ in the AF phase (below $T_N \approx 370$ °C). Hence, the primary objective of this study is to investigate the local fields at the Bi site below the Néel temperature, utilizing $^{111m}$Cd as a tracer ion.

This work combines time differential perturbed angular correlation (TDPAC) spectroscopy, X-ray diffraction (XRD), scanning electron microscopy (SEM), energy dispersive X-ray spectroscopy (EDXS), and \textit{ab-initio} density functional theory (DFT) to fully understand the complex characteristics of BFO.

\section*{Experimental details}
\subsection*{Sample preparation and characterization}

Polycrystalline BFO ceramic samples were synthesized via the solid-state reaction. 0.005 mol high-purity Bi$_2$O$_3$ (99.9\%, Acros Organics) and 0.005 mol high-purity Fe$_2$O$_3$ (99.9\%, Alfa Aesar) powders were weighed and thoroughly mixed through ball-milling for $\sim$ 2 days at 10 rotations per minute (rpm), using ZrO$_2$ balls and ethanol as a medium. The mixture was dried in air and then calcined at 820 $^\circ$C in air for 3 hours in an alumina crucible. The calcined powders were manually milled for 45 minutes with ethanol and then dried in air. A mixture of calcined powders and 3\% polyvinyl alcohol (PVA) was pressed at 5000 N ($\approx$ 100 MPa) to fabricate pellets with a diameter of 8 mm and a thickness of 1.5 mm. The pellets were then covered with calcined powders (using a powder bag), mounted in an alumina crucible, and finally sintered in air at 820 $^\circ$C for 6 hours.

To identify the primary phase (BiFeO$_3$) and the secondary phases of the synthesized samples, X-ray diffraction (XRD) measurements were conducted before and after the TDPAC measurements, using a PANalytical X-ray diffractometer, with the X’Pert PRO model attached with the X’Celerator detector using Cu-K$\alpha$ radiation ($\lambda_{\alpha1} = 0.154060$ nm, $\lambda_{\alpha2} = 0.154443$ nm) at 40 kV and 40 mA. A step angle of 0.01313$^\circ$ was used for data acquisition. Two BFO samples were studied, one sample before TDPAC measurement (sample ID: BFO2c) and another sample after TDPAC measurement (sample ID: BFO2b). They were cut from the same BFO pellet.

The grain size, atomic composition of the ceramics and the homogeneity of the elements were evaluated by scanning electron microscopy (SEM) and energy dispersive X–ray spectroscopy (EDXS) using an analytical SEM device (ESEM Quanta 400 FEG, FEI company) with an energy resolution $<$ 132 eV for Mn-K$\alpha$ radiation and a detector area of 10 mm$^2$. A high voltage of 30 kV was used to accelerate the electron beams. Two BFO samples (sample IDs: BFO2b and BFO2c) were studied. The experimental processes are summarized in Table I in the supplementary material (SM).

\subsection*{Ion implantation and TDPAC methodology }

The bulk BFO samples were implanted at room temperature with $^{111m}$Cd at the General-Purpose Separator (GPS) low mass (GLM) beam line, located at the Ion Separator OnLine DEvice-European Organization for Nuclear Research (ISOLDE-CERN) \cite{Catherall2017}. First, the molten Sn target at the GPS was bombarded with a 1.4 GeV proton beam. After ionization using the VADIS source \cite{Penescu2009}, the resulting ion beam was separated by mass and the $^{111m}$Cd isotope was delivered to the GLM beam line \cite{Schell2020}. This pure beam of $^{111m}$Cd, with an energy of 30 keV, was then implanted into the BFO samples with a dose of about 3 $\times$ 10$^{11}$ atoms/cm$^2$. The ionic implantation depth was simulated using the SRIM software package \cite{Ziegler2008}, based on a density of 8.34 g/cm$^3$. The implantation depth distribution shows a maximum at around 12 nm (Fig. \ref{figS1}). After the $^{111m}$Cd implantation, the implanted samples were annealed at 800 $^\circ$C in air for 20 minutes. Then, the full TDPAC study was conducted at various temperatures ranging from 15 $^\circ$C to 850 $^\circ$C within the ISOLDE solid state physics laboratories \cite{Johnston2017}.

\begin{figure*}[!htb]
\centering
\includegraphics[width=13cm]{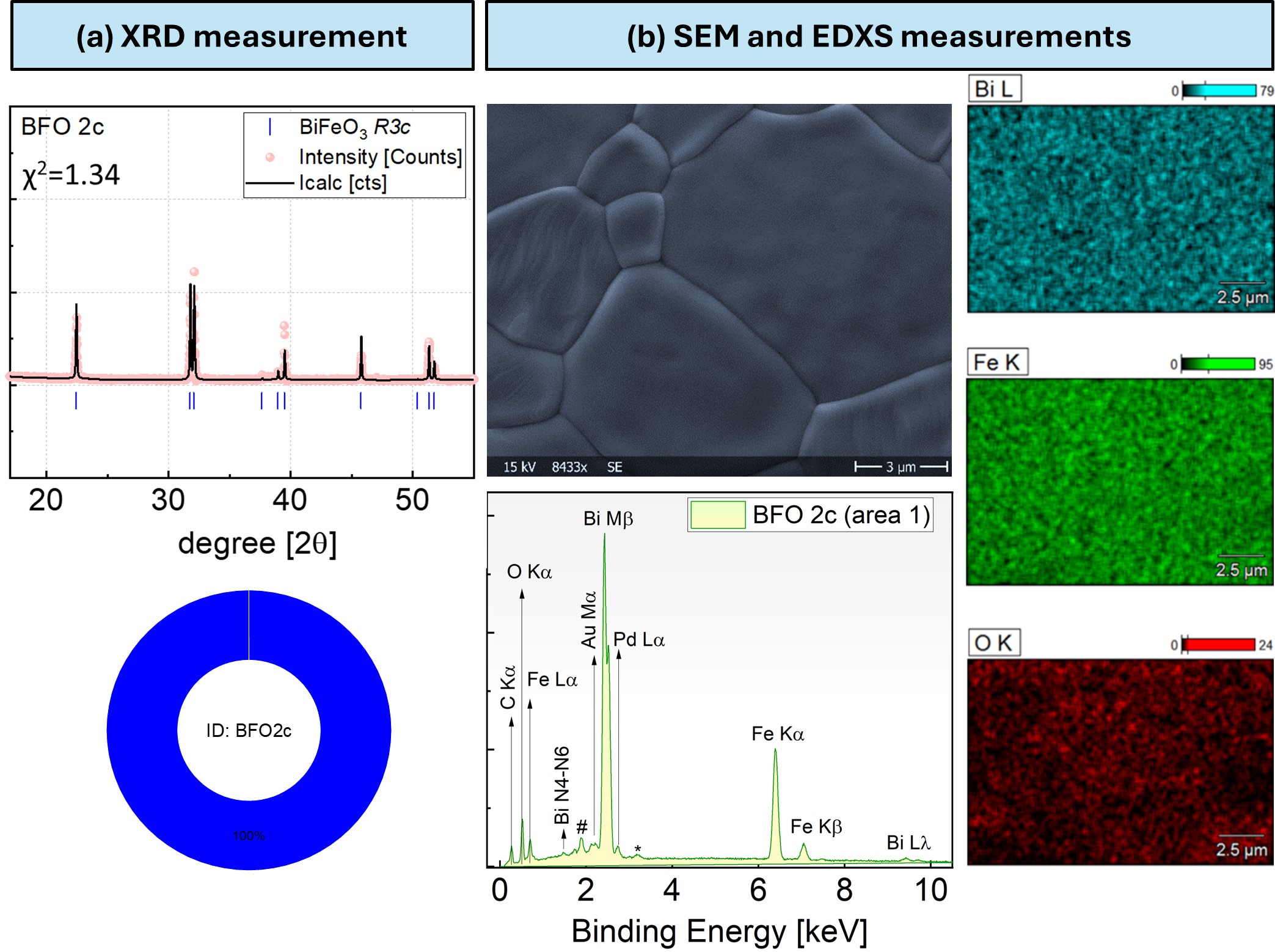}
\caption{(a) X-ray diffraction (XRD) diffractogram and the corresponding Rietveld refinements fit, with goodness of fit ($\chi^2$) values for the sample (ID: BFO2c) at room temperature. The corresponding pie charts show 100\% main phase BiFeO$_3$ (blue). The XRD diffractograms are plotted in terms of 2$\theta$, with a range of 15–55$^\circ$ for better visibility. (b) SEM image, EDXS spectrum and elemental mapping of sample BFO2c (fresh sample). The unassigned peaks (\#, *) originate from the sample holder and carbon residues of the carbon tape that was used to attach the sample to the sample holder.
}
\label{fig2}
\end{figure*}

The TDPAC method employs a trace amount of radioactive probe ($\sim 10^{11}$ atoms/cm$^2$) which is substituted into a well-defined chemical environment in a crystal by implantation, diffusion, or neutron activation \cite{Schatz1996, Abragam1953, Frauenfelder1966}. These isotopes, known as “probe atoms”, have specific nuclear properties. On the one hand, it is necessary for the probe atom to decay via a $\gamma$-$\gamma$ cascade. On the other hand, the intermediate level of this cascade must have sufficiently large electric quadrupole moment $Q$ and nuclear magnetic dipole moment $\mu$ (Fig. \ref{figS2}a) in order to interact with the local extranuclear electric field gradient (EFG) and magnetic field. In BFO, the EFG primarily originates from the spontaneous electric polarization induced by the displacement of the Bi$^{3+}$, while the magnetic field originates from the spins of the iron atoms \cite{Kubel1990}. Due to the interaction between the EFG (or magnetic field) and the quadrupole moment $Q$ (or magnetic dipole moment $\mu$), a precession (gyroscopic motion) of the nuclear spin occurs in the intermediate level (Fig. \ref{figS2}a). The gyroscopic movements can be seen as a modulation of the lifetime curves with the gyro frequencies (Fig. \ref{figS2}c-d, blue and orange colours).

The interaction between the EFG and the quadrupole moment $Q$ is called electric quadrupole interaction, and the gyro frequency is calculated by \cite{Schatz1996}:

\begin{equation}
\nu_Q = \frac{eQV_{zz}}{h},
\label{e1}
\end{equation}

where $e$ is the electron charge and $h$ is Planck's constant. For $^{111m}$Cd, $Q = +0.664(7)$b \cite{Haas2021}. $V_{zz}$ is the largest component of the traceless EFG tensor in the principal axis system. It relates to other main components and the asymmetry parameter of EFG$\eta$ by this relation \cite{Butz1987}:

\begin{equation}
\eta = \frac{V_{xx} - V_{yy}}{V_{zz}}
\label{e2}
\end{equation}

The interaction between the magnetic field $B$ and the magnetic dipole moment $\mu$ is called magnetic hyperfine interaction, and the gyro frequency is calculated by \cite{Schatz1996}:

\begin{equation}
\omega_L = -\frac{\mu}{I} B
\label{e3}
\end{equation}

where $I$ is the nuclear spin of the intermediate state. For $^{111\text{m}}$Cd, $I = \frac{5}{2}$ and $\mu \left( \frac{5}{2}^+ \right) = -0.7656(25) \, \mu_N$. Here, $\hslash$ is the reduced Planck's constant.

The frequencies in equations \ref{e1} and \ref{e3} are derived from the perturbation function (Eq. s4, SM) that was used to fit TDPAC spectra. The simulated TDPAC spectra using Eq. s4 (SM) for pure electric quadrupole interaction are presented in Fig. \ref {figS3} (SM).

\section*{Results and discussions }
\subsection*{XRD, SEM and EDXS}

Figure 2 shows the experimental results of XRD, SEM and EDXS. SEM indicates that the samples possess a polycrystalline structure, with a grain size in the $\mu$m range. Additionally, the XRD and EDXS analyses confirm that the synthesized samples are of high purity BiFeO$_3$ and have a homogeneous distribution of each element. The ratio of Bi and Fe atoms is almost 1:1. The XRD and EDXS analyses for the sample after TDPAC measurements indicate that the TDPAC process (including ionic implantation, thermal treatments, and TDPAC measurements) does not result in any significant alteration to the main phase and structure of BFO. The detailed results and discussions are reported in sections III and IV in the SM.

\section*{First-principles density functional theory calculations}

Density functional theory (DFT) calculations were performed with the Vienna ab-initio Simulation Package (VASP, v. 5.4.1) \cite{Kresse1996}, with the PAW method \cite{Blochl1994}. The recommended PAW potentials were used. Two supercells were constructed with a Cd atom replacing one Bi atom: one with 2$\times$2$\times$1 size with respect to the hexagonal cell parameters of the BFO R3c structure, with 120 atoms, and another with 3$\times$3$\times$1 size and 270 atoms. The calculations were spin-polarized, with the GGA+U approximation \cite{Perdew1996, Dudarev1998}, and a U value of 3 eV. The atomic forces were relaxed to more stable positions, while keeping the lattice parameters fixed in the optimized pure BFO lattice constants. An energy cutoff of 500 eV for the plane waves and a 4$\times$4$\times$3 k-points grid were used for the smaller supercell, and a 400 eV cutoff with a 2$\times$2$\times$1 k-points grid for the larger supercell. Different states were obtained with different magnetic moments depending on the starting point of the calculation, and in both cases the most stable solutions are found when using the wave functions obtained starting from higher U values and decreasing U until U=3 eV. The structure was relaxed until the norms of the atomic forces were all smaller than 0.05 eV/Å.

The electric field gradient (EFG) at Cd is converging with supercell size, with values $V_{zz} = 55.6$ and $57.7$ V/Å$^2$ for 2$\times$2$\times$1 and 3$\times$3$\times$1 supercells, respectively, with $\eta=0$ in both cases. However, while the 2$\times$2$\times$1 supercell results in a small total magnetic moment of 0.14 $\mu_B$ in the whole unit cell, this result is sensitive to the supercell size, as the calculations with the 3×3×1 supercell show a magnetic moment of 2.53 $\mu_B$. The prohibitive computational cost prevents us from calculating with larger supercells, but the present results strongly indicate the moment is not close to zero for a highly diluted impurity, as first suggested by the 2$\times$2$\times$1 supercell. The uncompensated moment comes mainly from the different local moment of Fe nearest to Cd, with absolute value of 1.97 $\mu_B$ instead of 4.06-4.07 like the other Fe neighbours, as shown in Table I. On the other hand, the moments are compensated in the smaller supercell calculation. The nearest Fe also slightly approaches Cd when going to the more diluted case, while the other Cd-Fe distances remain the same.

\begin{table}[h]
\centering
\caption{DFT local magnetic moments ($\mu_B$) for Fe at the smallest distances to the Cd impurity (Å), comparison between the two different supercells used.}
\begin{tabular}{|c|c|c|c|}
\hline
\multicolumn{2}{|c|}{\textbf{2$\times$2$\times$1 supercell}} & \multicolumn{2}{|c|}{\textbf{3$\times$3$\times$1 supercell}} \\
\hline
Fe-Cd distance & Fe magnetic moment & Fe-Cd distance & Fe magnetic moment \\
\hline
3.11 & -4.05 & 3.07 & 1.97 \\
3.28 (3 atoms) & 4.04 & 3.28 (3 atoms) & -4.07 \\
3.49 (3 atoms) & -4.05 & 3.48 (3 atoms) & 4.06 \\
3.74 & 4.05 & 3.74 & -4.06 \\
\hline
\end{tabular}
\end{table}

\begin{figure*}[!htb]
\centering
\includegraphics[width=13cm]{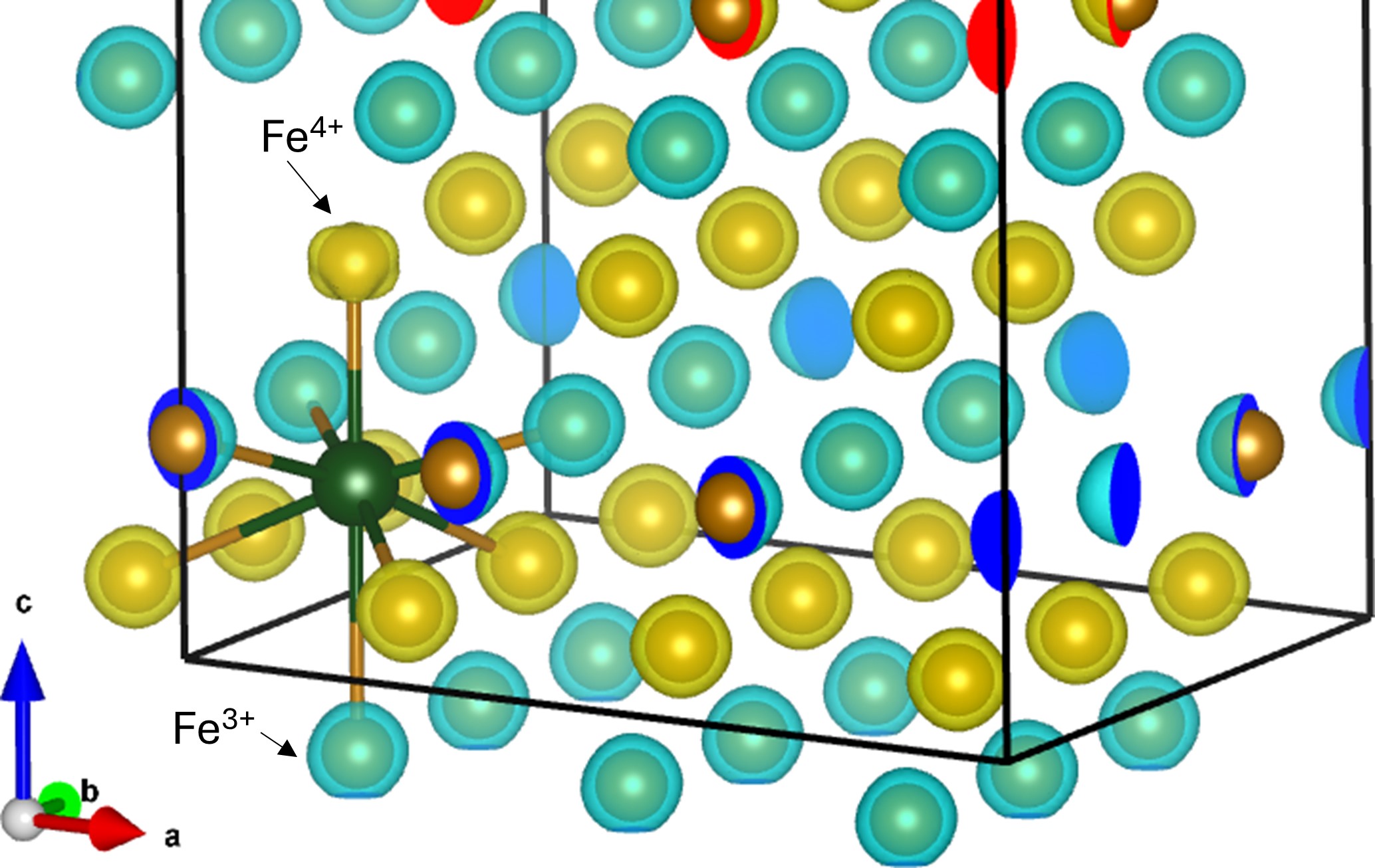}
\caption{ Spin difference density isosurface. The Cd atom is dark green and bonded to its nearest Fe atoms (brown). The structure is shown in the hexagonal setting (c is the polar direction). Blue and yellow surfaces represent opposite spin densities. For clarity, Bi and O atoms are not shown.
}
\label{fig3}
\end{figure*}

Fig.\ref {fig3} shows isosurfaces of the spin difference density, showing opposite spin densities in Fe, and no spin difference at Cd. It is noteworthy that the spin density in the Fe nearest to Cd has different shape.

In short, first-principles density functional theory calculations were performed with Cd replacing Bi in a BFO 3$\times$3$\times$1 supercell, including comparison with a 2$\times$2$\times$1 supercell. The magnetic moments of the Fe atoms neighbouring Cd are shown in Table I. In the 3$\times$3$\times$1 supercell, after structural relaxation, the Fe atom nearest to Cd that does not have a compensating magnetic moment, being smaller than the others, with 1.97 $\mu_B$ could be Fe$^{4+}$ (Fig. 3). However, the Fe$^{4+}$ was not identified in the 2$\times$2$\times$1 supercell because the Fe atom nearest to Cd has a compensating magnetic moment. The total magnetic moment calculated at Cd is zero for both 2$\times$2$\times$1 and 3$\times$3$\times$1 supercells. This implies that the effective magnetic field at Cd is always zero. The calculated EFG ($V_{zz} \approx 58$ V/Å$^2$, $\eta=0$) is in good agreement with experimental observations in terms of local EFG ($V_{zz}$), charge distribution around the probe ($\eta$), and balanced charge state (Fe$^{4+}$), which will be discussed in detail in the next section.

\section*{TDPAC results and discussions}

The TDPAC spectra $R(t)$ were fitted using the theoretical function (s4) presented in the supplementary material. The results are shown in Fig.\ref{fig4} These spectra are consistent with the simulated $R(t)$ spectra for pure electric quadrupole interaction using TDPAC theory as presented in Fig. \ref{figS3} in the supplementary material (corresponding to $\eta = 0$). This indicates that no combined magnetic and quadrupole interaction was observed at temperatures as low as 15 °C. The sole interaction observed in this study is that of purely electric quadrupole nature. The experimental data are in good agreement with the theoretical function (s4) when the damping factor ($\delta$) and the asymmetry parameter ($\eta$) are set to zero. This is in accordance with the results of DFT calculations. This reflects the precise substitution of the $^{111m}$Cd probe on the lattice site and that the electric charge distribution around the probe is symmetrical. This indicates that in the vicinity of the Cd probe, atomic vacancies or interstitial atoms are not formed, and the local structure remains undistorted under the substitutional location of the probe atom on the lattice site. The electric quadrupole frequencies $\nu_Q$ are converted into the main component $V_{zz}$ of EFG using the equation (\ref{e1}), and the temperature dependence of $V_{zz}$ and $\nu_Q$ are displayed in Fig.\ref{fig5}. The values of $V_{zz}$ are almost unchanged ($\approx 57$ V/Å$^2$) over the investigated temperatures, which is close to the calculated value ($V_{zz} \approx 58$ V/Å$^2$) obtained by DFT for the Cd substitutional on the Bi site. At 850 °C, the observed electrical quadrupole frequency increases to 112 MHz with highly asymmetrical charge distribution ($\eta = 0.86$) and no damping ($\delta = 0$). These values correspond to $\beta$-phase BFO, as reported by TDPAC measurements and DFT calculations in the work of Marschick et al. \cite{Marschick2020}. The experiments above the Néel temperature (400 °C < T < 800 °C) are not conducted in this study, because they were investigated in the earlier work \cite{Marschick2020}. Here, the measurements at $T = 400$ °C and $T = 800$ °C were intentionally performed to evaluate the $R(t)$ spectra in the paramagnetic phase (PM) and to compare with Marschick’s data. The $R(t)$ spectrum at 800 °C of this study is similar to that of Marschick’s data, but without damping. This reflects the fact that our samples are reproducible with higher material quality. In the present work, the experimental data were acquired using two spectrometers, KATAME \cite{Nagl2010} and K1 \cite{Schell2017}, to guarantee a high precision of the experimental results.

\begin{figure*}[!htb]
\centering
\includegraphics[width=8cm]{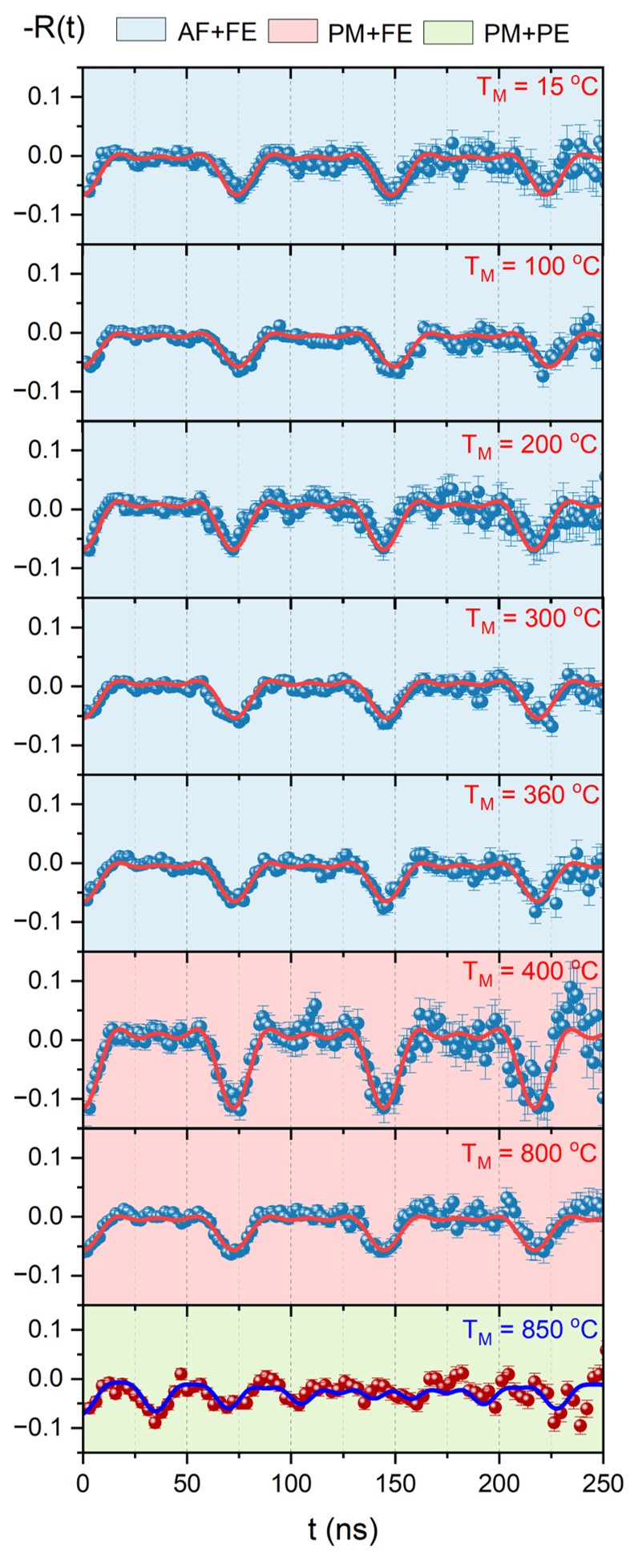}
\caption{The temperature-dependent TDPAC spectra $R(t)$ were fitted using the PACFit software \cite{Cavalcante342}. The spectra at 15 °C, 200 °C, 400 °C, and 850 °C were obtained using the KATAME \cite{Nagl2010} spectrometer whereas other spectra were measured at K1 \cite{Schell2017}. Blue, red, and green backgrounds denote spectra in the antiferromagnetic and ferroelectric (AF + FE), paramagnetic and ferroelectric (PM + FE), and paramagnetic and paraelectric (PM + PE) temperature ranges, respectively.}
\label{fig4}
\end{figure*}

\begin{figure*}[!htb]
\centering
\includegraphics[width=13cm]{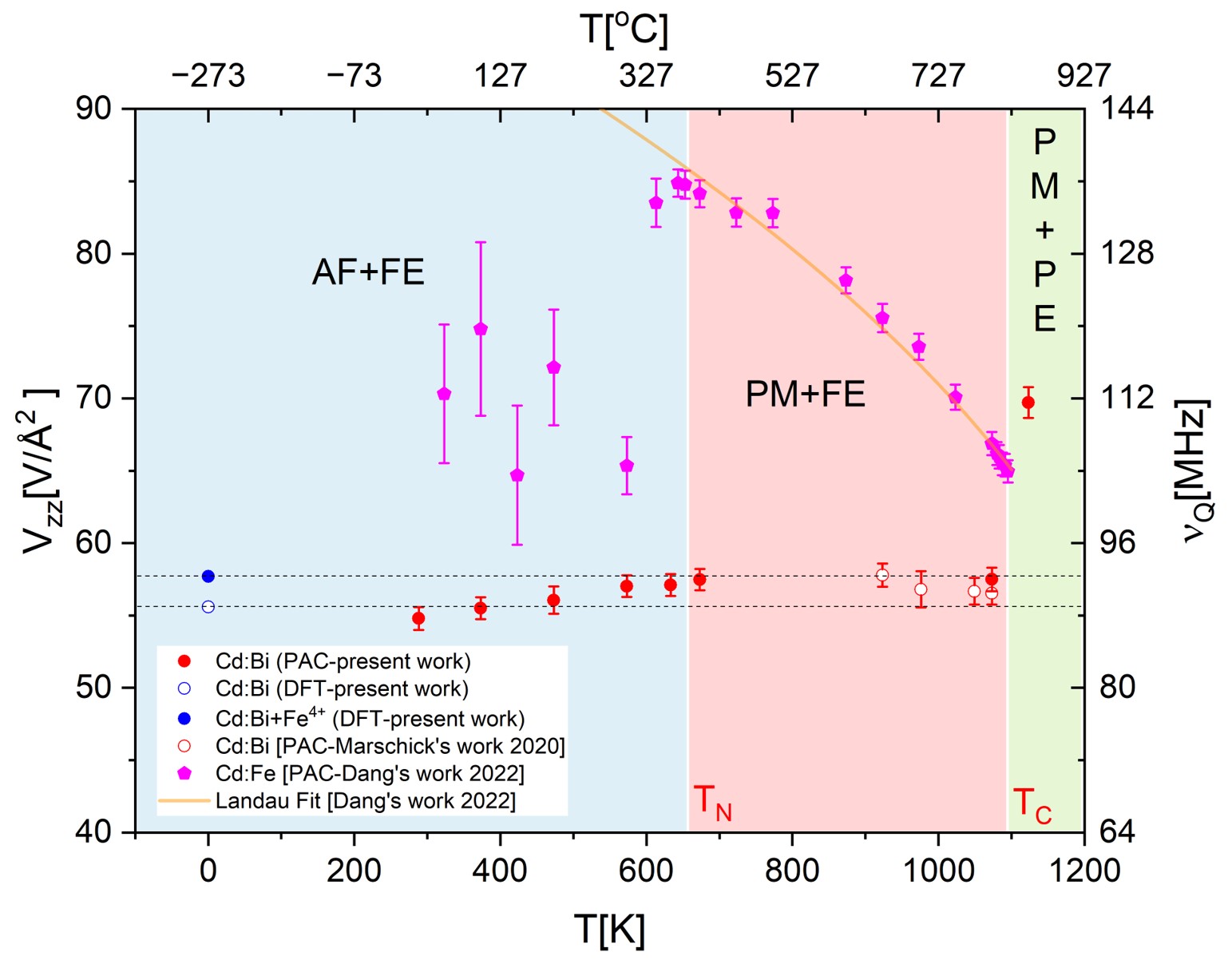}
\caption{Temperature dependence of $V_{zz}$ corresponding to the Cd-substitutional Bi site. (AF: Antiferromagnetic, FE: Ferroelectric, PM: Paramagnetic, PE: Paraelectric). $\nu_Q$ was extracted from the fittings of $R(t)$ shown in Fig. \ref{fig4} and $V_{zz}$ was converted from $\nu_Q$ according to equation (\ref{e1}). The data obtained by Marschick et al. \cite{Marschick2020} at high temperatures are compared to the present data in order to fully evaluate the ferroelectric order at the Bi site. The data of Dang et al. \cite{Dang2022} for the Cd-substitutional Fe site are incorporated into the present data in order to enable a comprehensive analysis of the ferroelectric order of BFO. The ferroelectric order at the Bi site of BFO follows a first-order phase transition, whereas the combined FE-AFD order at the Fe site follows a first-order close to second-order phase transition \cite{Strukov1998}. The Landau fit (see Eq. (s6) in the SM) was applied to the combined FE-AFD order at the Fe site of BFO \cite{Dang2022}. Below $T_N$, magnetoelectric coupling of the magnetostrictive origin suppresses the electric polarization at the Fe site \cite{Lee2013}. The two dashed lines serve to highlight the DFT results.
}
\label{fig5}
\end{figure*}

The radioactive probe is sensitive to the structural distortions of its direct and next-nearest neighbourhoods. Consequently, the EFG observed at Cd$^{2+}$-substitutional Bi site, as illustrated in Fig.\ref{fig1} , is primarily the result of the effective spontaneous electric polarization $P_s$ due to the displacements of the neighboring Bi atoms within the distorted rhombohedral unit cell. Furthermore, the antiferrodistortive (AFD) order, which co-exists with the ferroelectric order, allows the oxygen octahedra to tilt and rotate along the [111] direction in opposite directions in adjacent pseudo-cubic unit cells \cite{Ederer2005, Karpinsky2017}. The anti-rotation of the oxygen octahedra is always accompanied by the displacement of the iron atoms \cite{Anatoly2012, Lee2013}. Due to the symmetrical restrictions (namely the presence of the 3-fold rotation axis in the rhombohedral lattice), the ionic shifts (Fe atoms) are directed strictly along the [111] direction, resulting in a secondary spontaneous electric polarization \cite{Karpinsky2017, Anatoly2012}. This secondary spontaneous electric polarization resulting from the AFD order insignificantly contributes to the total EFG at Cd$^{2+}$.

The temperature independence of the EFG at the Cd$^{2+}$-substitutional Bi site follows the first-order phase transition of the ferroelectric order of the Bi-site. As defined in literature \cite{Strukov1998}, under the conditions of complete thermodynamic equilibrium in the system, the spontaneous electric polarization is independent of temperature, and a phase transition would have occurred at a temperature $T_0$ corresponding to the condition of equality of the specific potentials. The definition of the temperature dependence of the order parameter for the first-order transition is presented in Fig. \ref{figS8} (SM). In BFO, the Bi atom stays in complete thermodynamic equilibrium within the investigated temperature range leading to the stable EFG up to 800 °C. It can be postulated that the Cd$^{2+}$-substitutional Bi site may stabilize the local structure in its immediate vicinity during thermal treatment due to the different ionic radius and charge state in comparison to the host lattice site. However, this argument may be flawed, because the radii of Cd$^{2+}$ and Bi$^{3+}$ were found to be similar, which are 0.95 Å and 1.03 Å, respectively \cite{Shannon1976}. The presence of different charge states induces defect polarization $P_d$ (as illustrated in Fig. \ref{fig1}), which is decoupled from the ferroelectric polarization $P_s$. Consequently, the existence of a Cd$^{2+}$-substitutional Bi site does not affect the FE order of BFO. This will be discussed in greater detail. The EFG exhibits a sudden increase at 850 °C (above $T_c$) (Fig. \ref{fig5}) due to the change in the coordination number from 6 (in R3c space group) to 8 (in Pbnm space group) of the Bi atoms. This phase transition was discussed in detail by Marschick et al. \cite{Marschick2020}.

As observed at the Bi site, a combination of the FE and AFD orders is present at the Fe site. Above the Néel temperature, $T_N$, the EFG at the Fe site, as reported in previous publications \cite{Schell2022, Dang2022}, exhibits a decrease in value with temperature and follows the Landau theory for a first-order close to second-order phase transition \cite{Strukov1998}. The detailed discussions for a first-order close to a second-order phase transition of the combined FE-AFD order at the Fe site are presented in the supplementary material (Sec. VI). It can be concluded from Fig. \ref{fig5} that the combined FE-AFD order of BFO follows a first-order phase transition at the Bi atom and follows a first-order close to second-order phase transition at the Fe ion position. These observations are consistent with those reported in BaTiO$_3$ \cite{Strukov1998}, where a first-order transition and a first-order close to second-order phase transition were also observed.

As discussed in the DFT calculations, Fe$^{4+}$ is found in the 3$\times$3$\times$1 supercell of BFO. The formation of the Fe$^{4+}$ ion is necessary to compensate the unbalanced charge due to Cd$^{2+}$ substituting Bi$^{3+}$, following Kröger-Vink notation \cite{Kroger1956}:
\begin{equation}
\text{CdO + BiFeO}_3 \rightarrow \text{Cd}_{\text{Bi}}^{'} + \text{Fe}_{\text{Fe}}^{\bullet} + \frac{1}{2} \text{Bi}_2 \text{O}_3 \uparrow + \frac{5}{2} \text{O}_0
\end{equation}

Moreover, according to DFT simulations, one of the atoms in the unit cell has a much different magnetic moment of 1.97 $\mu_B$ (Table I). This would be the Fe$^{4+}$, which is located nearest to Cd$^{2+}$ in the polar direction (c axis in hexagonal setting in Fig. \ref{fig3} or on [111] direction in rhombohedral unit cell in Fig. \ref{fig1}). Magnetic moments (absolute values) of the other five iron atoms (Fe$^{3+}$) have a magnitude of about 4.06 $\mu_B$. These values are compatible with neutron scattering measurements \cite{Lebeugle2008} indicating a magnetic moment of iron $\mu_{\text{Fe}} = 4.11(15) \mu_B$. Bi-site substitution with heterogeneously divalent elements therefore generates Fe$^{4+}$, which is in good agreement with previous findings in the literature \cite{Zhou2005, Goncalves2018}. The spatial arrangement of the Fe$^{4+}$ ion nearest to the Cd$^{2+}$ ion results in the formation of a defect polarization $P_d$ pointing from the Cd$^{2+}$ to the Fe$^{4+}$ (see Fig. \ref{fig1}). In particular, one defect Cd$^{2+}$ ion is located on the Bi$^{3+}$ site, exhibiting a negative charge (–e), while another direct neighbor defect Fe$^{4+}$ ion is situated on the Fe$^{3+}$ site, exhibiting a positive charge (+e). The two defects thus form a defect dipole. This defect dipole generates defect polarization $P_d$, which is found to be decoupled from the ferroelectric polarization $P_s$. This is confirmed by the DFT calculations presented in the previous section. DFT calculations were performed at 0 K for the Cd-substitutional Bi site with and without the Fe$^{4+}$ in direct vicinity. The observed electric field gradients (EFGs) for both cases were found to be nearly identical ($V_{zz} = 57.7$ and 55.6 V/Å$^2$, respectively). This indicates that the presence of Fe$^{4+}$ does not significantly alter the EFG at the Bi site. In other words, the defect polarization $P_d$ is decoupled from the spontaneous ferroelectric polarization $P_s$.

In general, in order to neutralize the system (BFO) following Cd$^{2+}$ substitution at the Bi site, bismuth and oxygen vacancies can be formed in addition to Fe$^{4+}$ given in Kröger-Vink notation \cite{Kroger1956}:
\begin{equation}
\text{CdO + BiFeO}_3 \rightarrow \text{Cd}_{\text{Bi}}^{'} + \text{Fe}_{\text{Fe}}^{'} + V_O^{\bullet\bullet} + \frac{1}{2} \text{Bi}_2 \text{O}_3 \uparrow + \frac{5}{2} \text{O}_0
\end{equation}
\begin{equation}
\text{CdO + BiFeO}_3 \rightarrow \text{Cd}_{\text{Bi}}^{'} + V_{\text{Bi}}^{'''} + \text{Fe}_{\text{Fe}}^{\times} + 2V_O^{\bullet\bullet} + \frac{1}{2} \text{Bi}_2 \text{O}_3 \uparrow + \frac{5}{2} \text{O}_0
\end{equation}

Nevertheless, the aforementioned vacancies result in the formation of a highly asymmetrical EFG ($\eta \neq 0$). This contradicts the experimental results and DFT calculations, which indicate that the asymmetry parameter ($\eta$) is always zero.

In our previous studies, which focused on the iron site of bismuth ferrite, we demonstrated that the magneto-electric coupling is highly pronounced in the iron-containing magnetic sublattice below $T_N$ \cite{Schell2022, Dang2022}. In these cases, the magnetic order induces a significant anti-rotation and tilt of the oxygen octahedra (AFD octahedral rotation), which subsequently results in further displacement of Fe atoms, thereby inducing an additional spontaneous electric polarization $\Delta \mathbf{P}_{\text{Fe}}$ at the Fe site due to exchange-striction \cite{Lee2013}. This polarization $\Delta \mathbf{P}_{\text{Fe}}$ couples with the Fe spins, contributing to the enormous magnetoelectric coupling between the AFM and AFD orders within the magnetic sub-lattice (Fe-site). This investigation is in good agreement with the theoretical approach that the nature of the magnetoelectric effect in BiFeO$_3$ lies in the reorientation of the AFD vector in either electric or magnetic field \cite{Popkov2016}. Remarkably, no such magnetoelectric effect is observed at the Bi site in the present study. The total spontaneous electric polarization at the Bi site is independent of temperature and unaffected by the magnetoelectric coupling between the AFM and AFD orders at the Fe site. This indicates that the AFD order has an insignificant contribution to the EFG at Cd$^{2+}$.

As one of iron sites is populated with Fe$^{4+}$ ion, the iron ions should have non-canceling magnetic moments despite antiparallel alignment. DFT calculations show a weak magnetization of the 3$\times$3$\times$1 supercell (2.53 $\mu_B$) with the presence of Fe$^{4+}$, which creates a magnetic field that may measurably couple to the magnetic moment of the probe. However, no effects of phase transition or any other effects of local magnetic fields are detectable at the Cd$^{2+}$-substitutional Bi site since the effective magnetic field at the Cd probe site (or Bi site) is zero (Fig. \ref{fig3}). The $^{111m}$Cd probe could not sense the weak magnetic field resulting from the weak magnetization of the unit cell (2.53 $\mu_B$) since this weak magnetic field does not reach the $^{111m}$Cd probe (at Bi site). These DFT calculations are in full agreement with the experimental results.

Bulk bismuth ferrite has been investigated to exhibit no net magneto-electric coupling on the macroscopic scale, which is typically attributed to the antiferromagnetic nature of the material and the cycloidal ordering of the Dzyaloshinskii-Moriya interaction-generated magnetic moments \cite{Dzialoshinskii1957, Moriya1960}. The spin-cycloid structure with a periodicity length of approximately 62 – 64 nm results in the cancellation of the net macroscopic magnetization and linear magnetoelectric coupling in bulk BiFeO$_3$ \cite{Catalan2009}. Nevertheless, our investigation has revealed that the magnetoelectric coupling of BFO is already absent at the unit cell level. Our investigation is in good agreement with high-resolution single crystal neutron diffraction techniques, which have confirmed that the emergence of antiferromagnetic order does not affect the electric polarization at the Bi site \cite{Lee2013}. It is noteworthy that the substantial magneto-electric coupling observed exclusively at the Fe site results from the interaction between the magnetic moments of the Fe atoms and the secondary spontaneous polarization induced by antiferrodistortion \cite{Schell2022, Dang2022}. The magnetic field induced by the iron spins does not couple with the pre-existing ferroelectric polarization $P_s$ at the Bi site (Fig. \ref{fig1} ), as concluded by PAC measurements (this study) and neutron diffraction measurements \cite{Lee2013}.

As previously stated, the AFD order induces lattice fluctuations, yet no net macroscopic distortion is observed \cite{Fechner2024}. Consequently, the characterisation of AFD order cannot be achieved through macroscopic approaches. The TDPAC technique enables the characterisation of the AFD order at the unit cell level. The TDPAC data indicates that the AFD order is more prominent at the Fe site than the Bi site. Consequently, the magnetoelectric effect resulting from AFD and AFM coupling is pronounced at the Fe site, whereas it is essentially absent at the Bi site. The TDPAC approach has enabled the ferroic order of each individual element of BiFeO$_3$ to be characterized independently, allowing for the identification of the specific contributions of each element to the overall ferroic order of BFO.

As the foreign probe $^{111m}$Cd might alter the supercell, care must be taken in interpreting the data, because a single probe ion Cd$^{2+}$ in such a supercell effectively corresponds to a high doping level. After meticulous consideration, it can be concluded that the Cd$^{2+}$ substitutional Bi site and its direct vicinity Fe$^{4+}$ do not disrupt the ferroelectric order and anti-ferromagnetic order of the host lattice (BFO). Firstly, Cd$^{2+}$-substitutional Bi site and its direct vicinity Fe$^{4+}$ do not destroy the ferroelectric order of BFO. As previously discussed, the presence of Fe$^{4+}$ does not affect the ferroelectric order at the Bi site, as the defect dipole and ferroelectric polarization are decoupled. Moreover, the ferroelectric order at the Bi atom follows a first-order phase transition. This verifies that the probe ion Cd$^{2+}$ does not disrupt the ferroelectric order of BFO. Secondly, the Cd$^{2+}$-substitutional Bi site and its direct vicinity Fe$^{4+}$ do not disrupt the anti-ferromagnetic order of the host lattice (BFO). The results of the density functional theory (DFT) calculations, presented in Fig. \ref{fig3}, indicate that Fe$^{4+}$ is present in the 3$\times$3$\times$1 supercell, with the smallest magnitude of the magnetic moment (1.97 $\mu_B$) among the other ions. This results in an uncompensated moment in the 3$\times$3$\times$1 supercell. Nevertheless, the induced magnetic moment from Fe at Cd$^{2+}$ is ultimately found to be zero. In the 2$\times$2$\times$1 unit cell, Fe$^{4+}$ is not present, because all irons exhibit the same magnetic moment magnitude (4.05 $\mu_B$). The magnetic moments are compensated, and thus in the end the induced magnetic moment from Fe at Cd$^{2+}$ is also zero. It can be concluded that the total magnetic field at the Cd$^{2+}$ substitutional Bi site is always zero, regardless of the presence of Fe$^{4+}$. The absence of a magnetic field at the Bi site can be attributed to the inability of the magnetic field from the Fe sites to reach the Cd$^{2+}$ site site or Bi$^{3+}$ site (when Cd is substituted at the Bi site, the distance between Fe and Bi, Fe and Cd is nearly identical due to the similar ionic radii of Bi$^{3+}$ and Cd$^{2+}$ ). Figure \ref{fig2} illustrates that Fe$^{4+}$ and Fe$^{3+}$ (on the [111] direction) retain opposite spins. This implies that the presence of Fe$^{4+}$ does not disrupt the AF phase of BFO, as the double exchange interaction between Fe$^{4+}$ and Fe$^{3+}$ (Fe$^{4+}$ – O – Fe$^{3+}$) has been known to enhance the ferromagnetic properties \cite{Marzouk2020}, which might give rise to anti-ferromagnetism \cite{Gennes1960, Palii2020}.

\section*{Conclusions}

Bismuth ferrite is a type-I multiferroic for which the electric and magnetic ordering temperatures differ considerably. The ordering parameters of ferroelectric order and magnetization thus act at distinctly different energies. Our data now show that this is due to the exclusiveness of the ordering processes to one of the two sublattices in this structure. The ferroelectric order is carried by the 6s electron lone pair ordering on the Bi-sub-lattice while the magnetic order acts on iron and oxygen only via super-exchange, both being established facts. The coupling in between both lattices is negligible according to our data. Thus, the much sought magnetoelectric coupling cannot be provided by the bulk crystal of BFO and, if ever existing, will rely on extrinsic coupling effects mediated through grain boundaries, surfaces, or dopants. The macroscopically vanishing magnetoelectric coupling is thus not due to the antiferromagnetic averaging of coupling or the complete annihilation of net magnetization due to the spin cycloid, but is already contained in the structure of the unit cell.


\noindent\textbf{Acknowledgements}\\
Financial support was provided by the Federal Ministry of Education and Research (BMBF) through Grants No. 05K16PGA, 05K22PGA, 05K22PGB and 05K19SI1, alongside support from the ISOLDE collaboration. Financial support from the Deutsche Forschungsgemeinschaft (DFG) through Project LU 729/21-1 (Project No. 396469149) is gratefully acknowledged. We would like to acknowledge the Portuguese Foundation for Science and Technology (FCT), Projects UIDB/04349/2020 and CERN/FIS-TEC/8680003/2021. We acknowledge the support of the European Union Horizon Europe Framework research and innovation programme under grant agreement no. 101057511 (EURO-LABS) and of the European Union Horizon 2020 Framework research and innovation program under grant agreement no. 654002 (ENSAR2). J. N. G. would like to acknowledge national funds through the FCT/MCTES (PIDDAC) for the project CICECO-Aveiro Institute of Materials, UIDB/50011/2020, UIDP/50011/2020 \& LA/P/0006/2020. We are thankful to Mr. Tobias Bochmann for technical support with the EDXS measurements. We appreciate PD. Vladimir V. Shvartsman for his contributions to data interpretation. We also acknowledge the support of all the technical teams at ISOLDE for their excellent work in delivering high-quality beams for TDPAC measurements and the team at the institute in Essen.\\

\noindent\textbf{Author contributions}\\
T. T. D. performed the $^{111m}$Cd implantation, the TDPAC measurements and fits, and the XRD measurements. T. T. D. also generated the artificial TDPAC signals, synthesized the new ceramic samples, and wrote the manuscript. J. S. contributed to the $^{111m}$Cd implantation, the TDPAC measurements, the writing of the manuscript, and the data analysis and interpretation. A. D. performed and analyzed the EDXS measurements, produced the Rietveld refinements for the XRD data, and contributed to scientific discussions. J. N. G. did DFT simulations. M. E. C. assisted with the synthesis of the new ceramic samples. D. L. supported the production of artificial TDPAC signals and data analysis. I. C. J. Y. supported the $^{111m}$Cd implantation and the TDPAC measurements and carried out language editing. A. M. G. supported the $^{111m}$Cd implantations. S. M. F. contributed to the $^{111m}$Cd implantation and SEM measurements. D. Z. contributed to the $^{111m}$Cd implantation and supported the TDPAC measurements. D. C. L. contributed to the writing of the manuscript, data analysis, and data interpretation.
\\

\section*{Supplementary material (SM)}

\subsection*{I. Simulation of distribution of stopping range of implanted $^{111m}$Cd isotope in BiFeO$_3$ (BFO) using SRIM}

\begin{figure}[!hb]
\setcounter{figure}{0}
\renewcommand{\thefigure}{S\arabic{figure}}
\centering
\includegraphics[width=8cm]{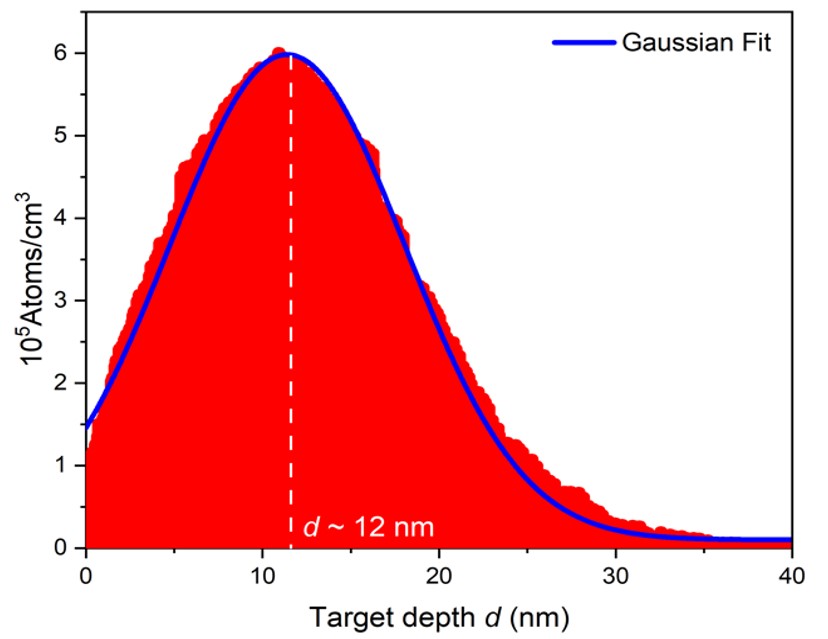}
\caption{Simulation of distribution of stopping range of implanted $^{111m}$Cd isotope in BFO using SRIM 2008 \cite{Ziegler2008}. The Gaussian fit (blue curve) shows the mean value about 12 nm. This means that the pure beam of $^{111m}$Cd, with an energy of 30 keV, after hitting the BFO target at the incident angle of about 0$^{\circ}$, will penetrate about 12 nm (d: penetrating depth) through the BFO target.}
\label{figS1}
\end{figure}

\subsection*{II. TDPAC technique}

\begin{figure}[!hb]
\renewcommand{\thefigure}{S\arabic{figure}}
\centering
\includegraphics[width=14cm]{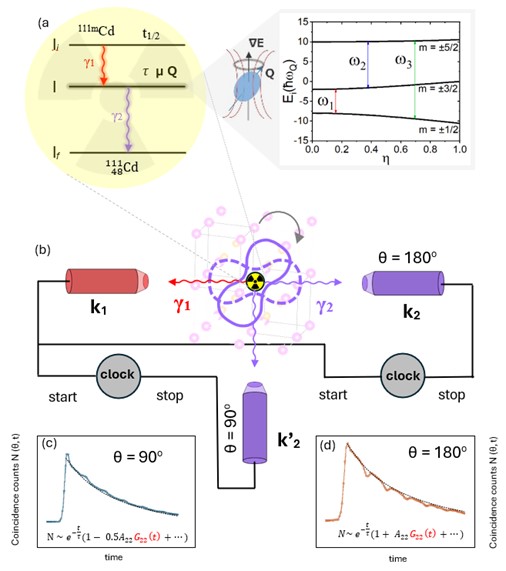}
\caption{(a) The decay scheme of $^{111m}$Cd \cite{Blachot2009}. The mother isotope $^{111m}$Cd, with the half-life time $t_{1/2} = 48.54$ minutes, decays to its ground state $^{111}_{48}$Cd via two gamma rays $\gamma_1$ (151 keV) and $\gamma_2$ (245 keV). The intermediate state has a spin $I = +5/2$ and the half-life $\tau = 84.5$ ns. The electric quadrupole hyperfine interaction splits the intermediate level into three degenerate sub-levels corresponding to different magnetic quantum numbers $m$ (from -5/2 to +5/2). The distance between two sub-levels is the transition frequency ($\omega_1, \omega_2, \omega_3$). The $m$-sublevel energies ($E_i$) change as a function of the asymmetry parameter ($\eta$). The electric quadrupole moment $Q = +0.83(13)$b \cite{Schatz1996} or $Q = +0.664(7)$b \cite{Haas2021} and the magnetic dipole moment is $\mu(5/2^+) = -0.7656(25) \mu_N$ \cite{Schatz1996}. (b) Representation of co-planar detectors in which each pair of detectors has an angle of 90$^\circ$ or 180$^\circ$ between them. The start detector ($k_1$) and stop detector ($k_2$ or $k_2’$) record the start signal ($\gamma_1$) and stop signal ($\gamma_2$), respectively. (c, d) The evolution of digitized coincidence counts with time corresponding to the 90$^\circ$ and 180$^\circ$ configurations of the start and stop detector.}
\label{figS2}
\end{figure}

Detailed descriptions of the TDPAC technique and its formalism can be found in literature \cite{Schatz1996, Abragam1953, Frauenfelder1966, Butz1989}. A simplified description of the TDPAC method can be found in the study of Catchen \cite{Catchen1995}. This manuscript presents the basic idea of TDPAC and its formalism in a (hopefully) understandable manner.

As mentioned in the main manuscript, the “probe atoms” must decay to their ground state via a cascade of two gamma rays as shown in Fig. \ref{figS2}a. From the excited state with spin $I_i$, the atomic nucleus passes into the intermediate state with spin $I$ by emitting a gamma quantum (in red). From this state, it then enters the ground state $f$ of the cascade by emitting a second gamma quantum (in purple). After the emission of the first $\gamma$ quantum, the nucleus of the probe atom is in the intermediate state of the cascade. Now, for reasons of spin conservation, there is a spatial probability distribution for the emission direction of the second gamma quantum with respect to the emission direction of the first gamma. There are, therefore, different probabilities for different angles between the emission directions of $\gamma_1$ and $\gamma_2$. In Fig. \ref{figS2}b, the first gamma quantum $\gamma_1$ has been emitted in the direction of detector $k_1$. The probability in which direction the second gamma quantum $\gamma_2$ is emitted is indicated by the emission lobe (in purple). In the case that the $\gamma_2$ is emitted parallel or antiparallel to the $\gamma_1$, there is a much greater probability than in the case that the $\gamma_2$ emission direction is perpendicular to the $\gamma_1$ emission direction (anisotropy).

In PAC experiments we have a set of detectors whose apparatus is usually done in such a way that each pair of detectors has an angle of 90$^\circ$ or 180$^\circ$ between them (Fig. \ref{figS2}b). If a certain gamma quantum $\gamma_1$ is detected by a detector $k_1$ with the energy $E_{\gamma1}$ the clock starts counting until another gamma quantum $\gamma_2$ is detected by a detector $k_2$ (or $k'_2$) with the energy $E_{\gamma2}$, making the clock stop. It is noticed that two gamma quanta must emit from the same single nucleus. Then one event from one nucleus will be stored in the coincidence counting rate of the pair of detectors. However, at time $t$, there are $N$ nuclei emitting gamma quanta simultaneously, resulting in $N$ recorded events (number of coincidence events). The number of coincidence events ($N$) decreases exponentially over time due to the decay of the intermediate state.

If no EFG (or magnetic field) acts on the spin of the probe atom while it is in the intermediate state, this emission lobe remains stable over time (Fig. \ref{figS2}b, purple dashed-line lobe). In this case, it is the undisturbed angular correlation. The number of coincidence events $N$ exponentially decreases over time (Fig. \ref{figS2}c-d, the exponentially dashed curves).

Now, however, an EFG (or magnetic field) acts on the spin of the probe atom while it is in the intermediate state. Due to the interaction between the EFG (or magnetic field) and the quadrupole moment $Q$ (or magnetic dipole moment $\mu$), a precession (gyroscopic motion) of the nuclear spin occurs at the intermediate level (Fig. \ref{figS2}a) and thus the emission lobe (purple lobe) is also rotated over time, it gyrates (Fig. \ref{figS2}b). In this case, it is the disturbed angular correlation, or PAC. The gyroscopic movements can be seen as a modulation of the lifetime curves with the gyro frequencies (Fig. \ref{figS2}c-d, blue and orange colors).

Now, we construct the experimental perturbation function \cite{Danielsen1995}:
\begin{equation}
R(t) = \frac{2 \left( N(180^\circ,t) - N(90^\circ,t) \right)}{N(180^\circ,t) + 2 \overline{N}(90^\circ,t)} \tag{s1}
\end{equation}

Under these hyperfine interactions (electric quadrupole and magnetic interaction), the number of coincidence events is modulated over time following this formula \cite{Butz1996}:
\begin{equation}
N(\theta, t) \sim e^{-t/\tau} \left(1 + A_{22} G_{22}(t) P_2(\cos\theta)\right) \tag{s2}
\end{equation}
where:
\begin{itemize}
  \item \( P_2(\cos\theta) \) is the Legendre polynomial for \( n = 2 \), with \( P_2(\cos(180^\circ)) = 1 \) and \( P_2(\cos(90^\circ)) = -1/2 \).
  \item \( A_{22} \) is the experimental anisotropy coefficient when finite detector and sample size are considered. Its value is typically reduced by 5\% to 15\% compared to the theoretical one \cite{Haas2021}.
  \item \( G_{22}(t) \) is the perturbation factor that will be discussed thoroughly later in this section.
  \item \( \tau \) is the half-life of the intermediate state.
\end{itemize}

Inserting (s2) into (s1), we obtain:
\begin{equation}
R(t) \approx A_{22} G_{22}(t) \tag{s3}
\end{equation}

Theoretically, the perturbation factor \( G_{22}(t) \) has the following form \cite{Dang2022}:
\begin{equation}
G_{22}(t) = \sum_{i=1}^{m} f_i \left[ a_{0i} + \sum_{n=1}^{30} a_{ni} \cos(\omega_{ni} t) \exp \left(-0.5 \left( (\delta_i \omega_{ni} t)^p + (\omega_{ni} \tau_R)^2 \right) \right) \right] \tag{s4}
\end{equation}

In Eq. (s4):
\begin{itemize}
  \item \( f_i \) is the fraction of probe atoms exposed to the hyperfine interaction \( i \) (i = 1, 2, 3, …), with \( \sum_{i=1}^{m} f_i = 1 \). Each “i” represents one local environment in a crystal.
  \item The \( \omega_{ni} \) are the transition frequencies taken from the 30 non-diagonal elements of the 6×6 frequency matrix (for spin \( I = 5/2 \)).
  \item The \( a_{ni} \) are taken from the 30 non-diagonal elements of a 6×6 amplitude matrix and are the amplitudes of the 30 corresponding transition frequencies; \( a_{0i} \) is determined from the sum of the 6 diagonal elements of the amplitude matrix and is called the “hard-core” value. The amplitudes \( a_{ni} \) and transition frequencies \( \omega_{ni} \) can be derived from the Hamiltonian matrix for spin \( I = 5/2 \) \cite{Dang2022}.
\end{itemize}

The asymmetry parameter \( \eta \) is not explicitly shown in Eq. (s4). It, however, can be extracted from the fitting software \cite{Cavalcante342}, and is defined as follows \cite{Butz1987}:
\begin{equation}
\eta = \frac{V_{xx} - V_{yy}}{V_{zz}}
\tag{s5}
\end{equation}

The \( V_{ij} \) (i,j = x, y, z) are the 3 main components of the EFG tensor in the principal axis system (PAS), which is defined as \cite{Autschbach2010}:

\begin{equation}
V^{\text{PAS}} = \begin{bmatrix}
V_{xx} & 0 & 0 \\
0 & V_{yy} & 0 \\
0 & 0 & V_{zz}
\end{bmatrix}
\tag{s6}
\end{equation}
\( V_{ij} \) must meet two conditions:
\\
V$_{xx}$ + V$_{yy}$ + V$_{zz}$ = 0 \quad \text{(The tensor is traceless due to the Laplace equation)}
\\
\\
$|V_{zz}| \geq |V_{yy}| \geq |V_{xx}|$ \quad \text{(in the principal axis system)}

Therefore, \( \eta \) is restricted to \( 0 \leq \eta \leq 1 \). With given \( \eta \) and \( V_{zz} \), the other two principal components are given by:

\begin{equation}
V_{xx} = -\frac{1}{2}(1 - \eta) V_{zz} \quad \text{and} \quad V_{yy} = -\frac{1}{2}(1 + \eta) V_{zz}
\tag{s7}
\end{equation}

Therefore, the electric field gradient tensor can be described for most applications by two parameters, i.e., \( V_{zz} \) and \( \eta \).

The asymmetry parameter \( \eta \) measures the symmetry of the field (charge distribution) with respect to the axial symmetry axis \( z \). When \( \eta = 0 \), \( V_{xx} = V_{yy} = -\frac{1}{2} V_{zz} \), the charge distribution is symmetrical with respect to the \( z \) axis. When \( \eta = 1 \), \( V_{xx} = 0 \), \( V_{yy} = -V_{zz} \), the charge distribution is highly asymmetrical with respect to the \( z \) axis.

The damping factor, \( \delta_i \), indicates the deviation of the transition frequency from the mean value. The values \( p = 1 \) and \( p = 2 \) in the exponent represent Lorentzian and Gaussian distributions, respectively \cite{Zacate2011}. \( \tau_R \) is the finite time resolution (FWHM) of the detectors.

We fitted the TDPAC spectra using Eq. (s4). From the fits, we can extract various hyperfine parameters such as the quadrupole interaction frequency \( \nu_Q \) caused by the electric quadrupole splitting, and the Larmor frequency \( \omega_L \) caused by the magnetic dipole splitting of the intermediate state. In addition, other parameters such as the asymmetry parameter (\( \eta \)), the frequency distribution (\( \delta \)), and the fraction (\( f \)), are extracted from the fit. These extracted parameters reveal information about the nature of the hyperfine interactions.

\begin{figure}[!htb]
\renewcommand{\thefigure}{S\arabic{figure}}
\centering
\includegraphics[width=12cm]{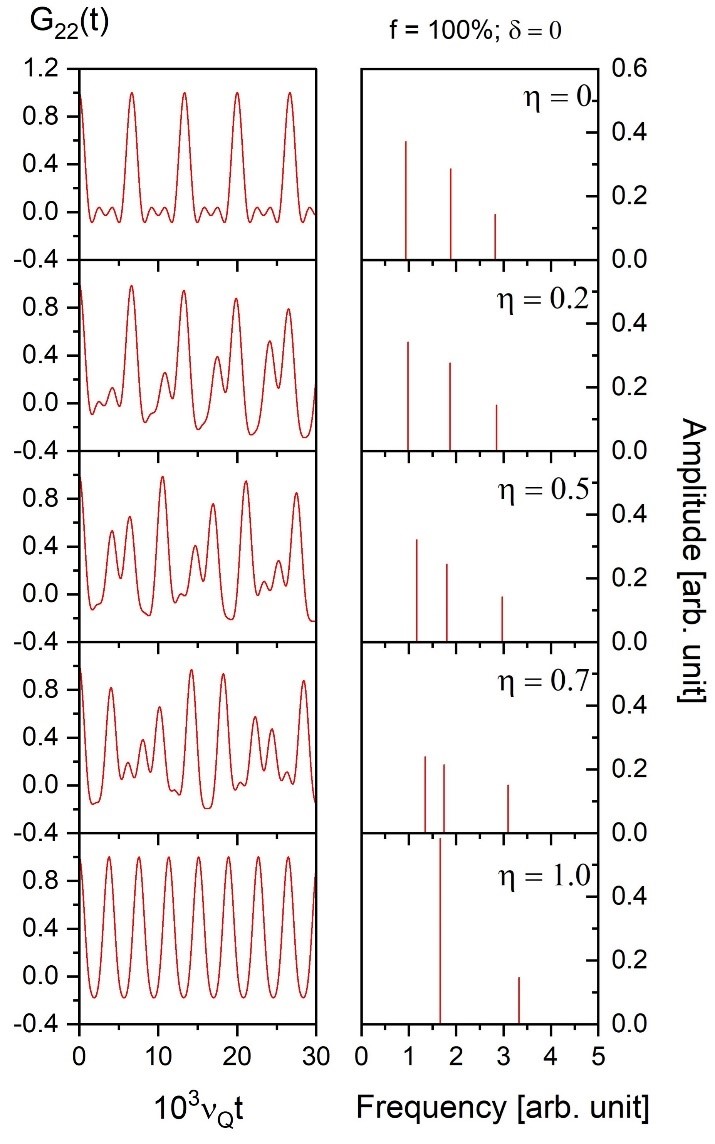}
\caption{(left) Simulated perturbation factor \(G_{22}(t)\) of the static electric quadrupole interaction for any value of quadrupole interaction frequency \(\nu_Q\), with one probe site (\(f = 100\%\)), with no damping (\(\delta = 0\)), and different asymmetry parameters \(\eta\) in the electric coordinate systems \cite{Bostrom1970}. (right) The corresponding Fast Fourier Transforms (FFTs) of the transition frequencies.}
\label{figS3}
\end{figure}

In this study, we also generate artificial TDPAC spectra for static electric quadrupole interactions based on TDPAC theory with the assistance of Python \cite{Python3.10.11} as shown in Figs. \ref{figS3}. These simulations are restricted to polycrystalline sources. Figure \ref{figS3} shows the simulated perturbation factor \( G_{22}(t) \) (or TDPAC spectra) and their corresponding Fast Fourier Transforms (FFTs) for static electric quadrupole interactions at one probe site, including neither damping (\( \delta = 0 \)) nor the temporal resolution of the detectors. Graphs of the simulated TDPAC spectra are plotted with increasing asymmetry parameter (\( \eta \)). It is noted that for the undamped electric interaction, a maximum of three and a minimum of two transition frequencies are visible in the FFTs due to the degree of degeneracy between the nuclear energy states of the spin \( I = 5/2^+ \) nucleus, as shown in the FFTs in Fig. \ref{figS2}. Moreover, we also observe that \( \omega_1 : \omega_2 : \omega_3 = 1:2:3 \) for \( \eta = 0 \) and \( 1:1:2 \) for \( \eta = 1 \).

\subsection*{III. Energy dispersive X–ray spectroscopy (EDXS) analysis }

\begin{figure}[!hb]
\renewcommand{\thefigure}{S\arabic{figure}}
\centering
\includegraphics[width=14cm]{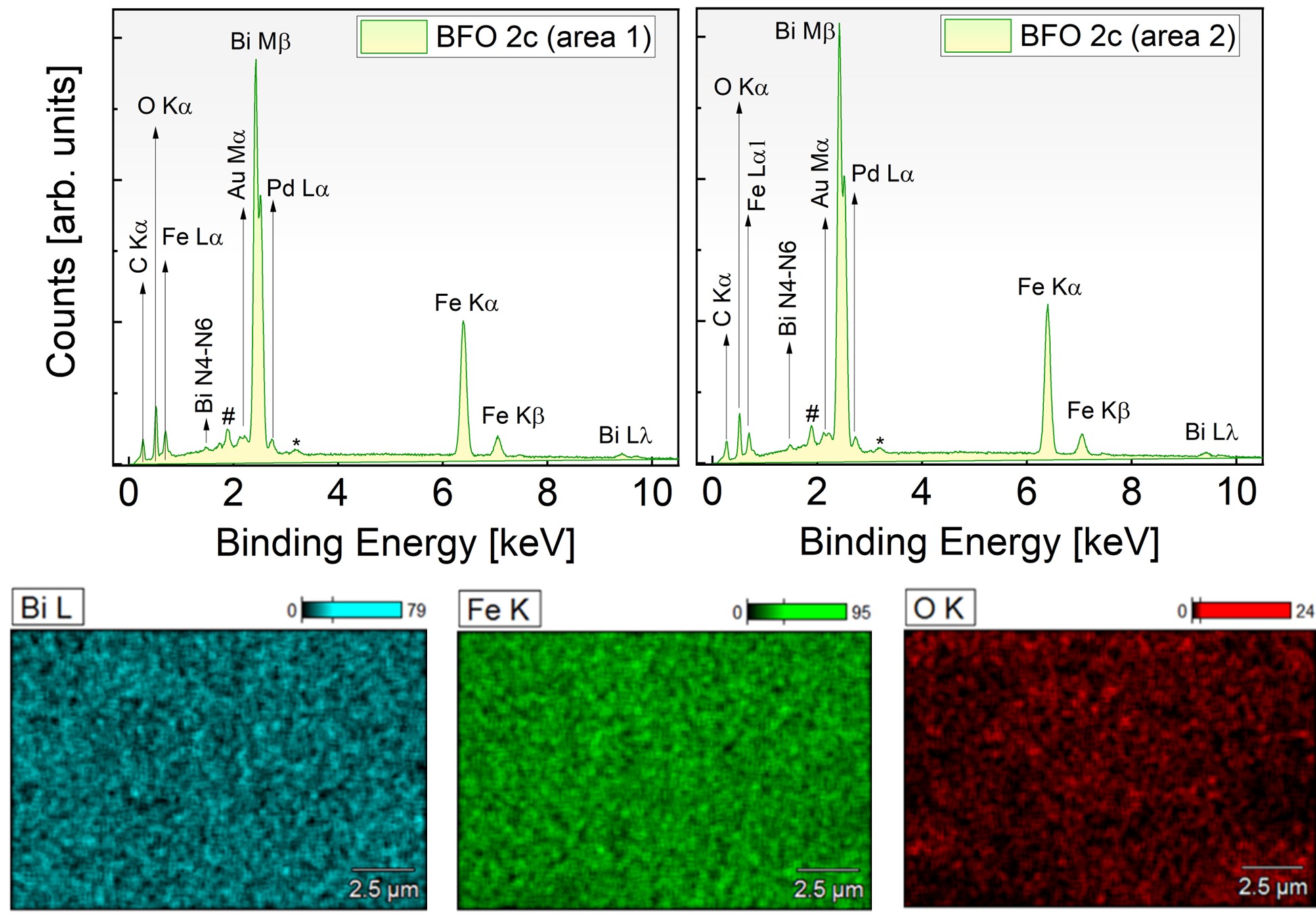}
\caption{The EDXS spectra and elemental mapping of sample BFO2c (fresh sample) are shown. The unassigned peaks (\#, *) might have originated from the sample holder and carbon residues of the carbon tape that was used to attach the sample to the sample holder.
}
\label{figS4}
\end{figure}

\begin{figure}[!hb]
\renewcommand{\thefigure}{S\arabic{figure}}
\centering
\includegraphics[width=14cm]{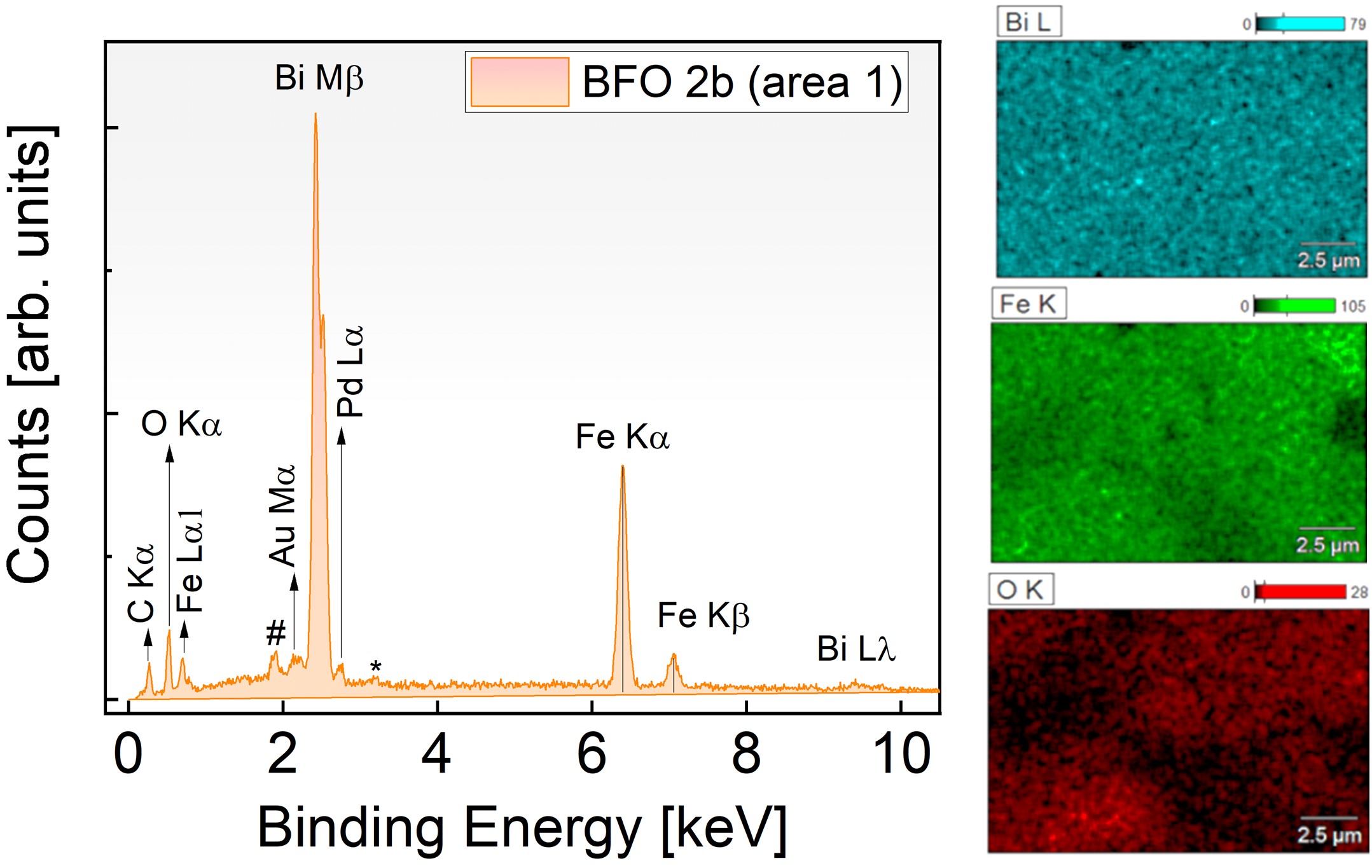}
\caption{The EDXS spectra and elemental mapping of sample BFO2b. The unassigned peaks (\#, *) might have originated from the sample holder and carbon residues of the carbon tape that was used to attach the sample to the sample holder.
}
\label{figS5}
\end{figure}

\begin{figure}[!htb]
\renewcommand{\thefigure}{S\arabic{figure}}
\centering
\includegraphics[width=12cm]{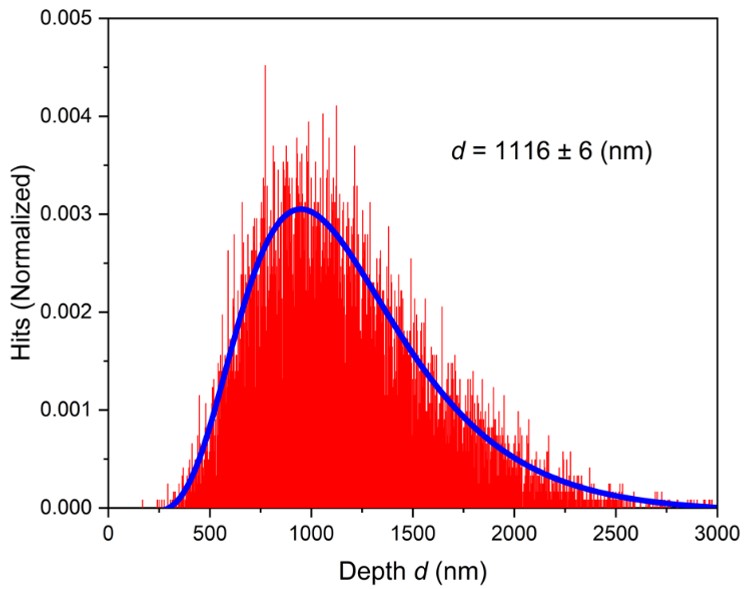}
\caption{A simulation of the electron penetration depth (EDXS measurement) in BFO ceramic using the CASINO software \cite{Alexandre2000}. The fitting curve (blue) shows a maximum at around 1 \(\mu\)m. This electron penetration depth is much deeper than the penetrating depth of the implanted \textsuperscript{111m}Cd (12 nm) as presented in Fig. \ref{figS1}.
}
\label{figS6}
\end{figure}

The experimental processes before the EDXS measurements are summarized in Table sI. EDXS measurements were performed to 1) confirm the formation of the BFO compound, 2) identify impurities, and 3) determine the homogeneity of the elements (up to 1 µm depth (Fig. \ref{figS6})). During the EDXS measurements, different areas of the sample were focused on, and the corresponding peaks are shown in Figures \ref{figS4} and \ref{figS5}.

\begin{table}[h]
\setcounter{table}{0}
\renewcommand{\thetable}{S\arabic{table}}
\centering
\caption{The experimental processes before the EDXS measurements.}
\begin{tabular}{|c|c|c|c|}
\hline
Sample ID & \textsuperscript{111m}Cd implantation & Thermal annealing & TDPAC measurement \\
\hline
BFO2c (Fresh) & --- & --- & --- \\
BFO2b & ~ 3.10\textsuperscript{11} atoms/cm\textsuperscript{2} & 800 °C in air for 20 minutes & Room temperature \\
\hline
\end{tabular}
\end{table}

Sample BFO2c has an almost 1:1 ratio of Bi and Fe atoms, as seen in Table II. EDXS mappings were done for two different areas (Fig. \ref{figS4}). The result implies that there is only a single phase of BFO. Bi-rich and Fe-rich impurities like selenite (Bi$_{25}$FeO$_{40}$) and mullite (Bi$_{2}$Fe$_{4}$O$_{9}$) are not present in this sample within the detection limit of EDXS. Each elemental mapping is displayed with a different relative intensity scale in which each element, denoted by color, is presented in bright (element-rich) or dark (element-deficient) contrast, as shown in Fig. \ref{figS4}. The synthesized R3c phase (rhombohedral $\alpha$-BFO in R3c setting) of BFO constitutes a uniform and homogeneous elemental distribution over the entire area that was measured for the sample, which confirms that the BFO pellet was well synthesized.

In sample BFO2b, the ratio of Bi and Fe is almost 1:1, as indicated in Table II. However, the concentration of Fe is marginally higher than that of Bi, which is possibly due to the formation of Fe-rich impurities (Bi\textsubscript{2}Fe\textsubscript{4}O\textsubscript{9} or Fe\textsubscript{2}O\textsubscript{3}) or the volatility of Bi at high temperature. The implanted Cd is under the detection limit of EDXS and could not be observed in the ceramic, as shown in Fig. \ref{figS5}. With an electron beam energy of 30 keV, the penetration depth of electrons inside the BFO ceramic is estimated to be around 1 µm (Fig. \ref{figS6}) using the monte CArlo SImulation of electroN trajectory in sOlids (CASINO) software (version 2) \cite{Alexandre2000}. This electron depth is much deeper than the Cd depth (~ 12 nm) as presented in Fig. \ref{figS1}. However, the probability of finding Cd at the specific area of the sample by using EDXS is infinitesimal because \textsuperscript{111m}Cd was aimed to be implanted into the sample with minimal concentration (~10\textsuperscript{11} atoms/cm\textsuperscript{2}) to avoid severe alteration to the structure of the sample. We also observe similar results in different areas of the sample. The elemental mapping in Fig. \ref{figS5} reveals a homogeneous distribution of each element, even though the sample underwent substantial ionic implantation and thermal treatment. On the mapping image for oxygen, several dark and shallow holes are observable. However, bright red spots are still visible inside these holes. This might be due to the sample surface not being flat. Therefore, within the resolution limit of the image map, we can conclude that the distribution of oxygen as an atomic element inside the BFO sample is homogeneous. Our result agrees with that of the unique technique used at ISOLDE-CERN \cite{Catherall2017, Schell2020}, in which case the implanted probe (\textsuperscript{111m}Cd) does not affect the structure of the BFO sample, which implies that the PAC process (including ionic implantation, thermal treatments, and PAC measurements) does not change the homogeneity of the element distribution.

\begin{table}[h]
\renewcommand{\thetable}{S\arabic{table}}
\centering
\caption{The elemental composition of the BFO samples.}
\begin{tabular}{|c|c|c|c|c|c|c|}
\hline
Sample ID & Element & Int. Cps/nA & Weight \% & Weight \% Error & Atom \% & Atom \% Error \\ 
\hline
BFO2c (area 1) & O & 104.16 & 17.3 & ± 0.3 & 63.0 & ± 1.0 \\ 
 & Fe & 767.52 & 18.1 & ± 0.1 & 18.7 & ± 0.3 \\ 
 & Bi & 678.93 & 64.6 & ± 0.8 & 18.3 & ± 0.2 \\ 
\hline
BFO2c (area 2) & O & 90.30 & 14.7 & ± 0.3 & 58.0 & ± 1.0 \\ 
 & Fe & 855.30 & 19.5 & ± 0.1 & 21.1 & ± 0.2 \\ 
 & Bi & 717.31 & 65.8 & ± 0.8 & 20.9 & ± 0.2 \\ 
\hline
Total (BFO2c) &  &  & 100.0 &  & 100.0 &  \\ 
\hline
BFO2b (area 1) & O & 74.80 & 13.6 & ± 0.6 & 56.3 & ± 2.6 \\ 
 & Fe & 751.06 & 18.8 & ± 0.4 & 22.2 & ± 0.4 \\ 
 & Bi & 678.29 & 67.6 & ± 1.7 & 21.5 & ± 0.5 \\ 
\hline
Total (BFO2b) &  &  & 100.0 &  & 100.0 &  \\ 
\hline
\end{tabular}
\end{table}

The formation of Fe-rich impurities (Bi\textsubscript{2}Fe\textsubscript{4}O\textsubscript{9} or Fe\textsubscript{2}O\textsubscript{3}) in sample BFO2b may arise from the thermal annealing processes conducted at high temperatures (800 \(^\circ\)C in air for 20 minutes). Marschick et al. demonstrated that BiFeO\textsubscript{3} begins to decompose into Bi\textsubscript{2}Fe\textsubscript{4}O\textsubscript{9} and Bi\textsubscript{25}FeO\textsubscript{39} at 700–720 \(^\circ\)C \cite{Marschick2020}. The exact phase of the specified impurities in these samples can be ascertained through XRD measurements, which will be covered in section IV.

\subsection*{IV. XRD measurements}

\begin{figure}[!htb]
\renewcommand{\thefigure}{S\arabic{figure}}
\centering
\includegraphics[width=10cm]{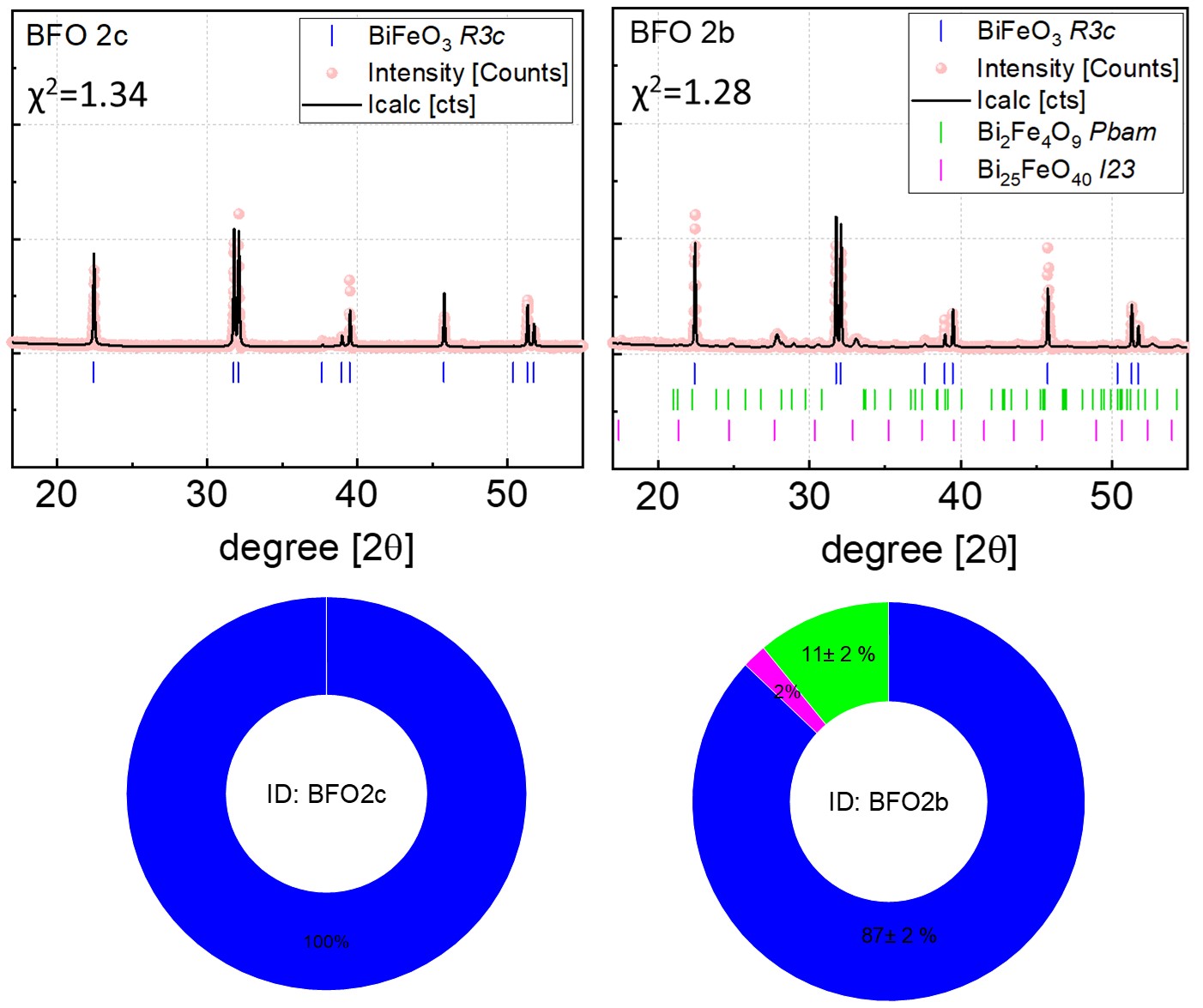}
\caption{X-ray diffraction (XRD) diffractograms and the corresponding Rietveld refinements fits, with goodness of fit (\(\chi^2\)) values for samples BFO2b, and BFO2c at room temperature. The corresponding pie charts show the proportion of main and secondary phases in each sample: BiFeO\(_3\) (blue), Bi\(_2\)Fe\(_4\)O\(_9\) (green), and Bi\(_{25}\)FeO\(_{40}\) (purple). The XRD diffractograms are plotted in terms of 2\(\theta\), with a range of 15–55\(^\circ\) for better visibility.
}
\label{figS7}
\end{figure}

Figure \ref{figS7} provides a detailed overview of the X-ray diffraction (XRD) diffractograms with their Rietveld refinement fits, accompanied by a pie chart showing the corresponding proportion of both the main and secondary phases in each sample. It has been determined that the Fe-rich impurity in question is the Bi\(_2\)Fe\(_4\)O\(_9\) Pbam phase, accounting for 12 $\%$ of the sample. The concentration of Bi\(_{25}\)FeO\(_{40}\) is within the detection limit (and hence the error interval) of 2 $\%$, and thus is practically negligible. As indicated in the work of Marschick et al. \cite{Marschick2020}, this little amount of Bi\(_2\)Fe\(_4\)O\(_9\) could not be detected by TDPAC because this phase is possibly not located in the first neighbors of Cd, and therefore it is not visible in \( R(t) \) spectra shown in Fig. \ref{figS3} in main manuscript.

\subsection*{V. Ferroelectric order} 

\begin{figure}[!htb]
\renewcommand{\thefigure}{S\arabic{figure}}
\centering
\includegraphics[width=12cm]{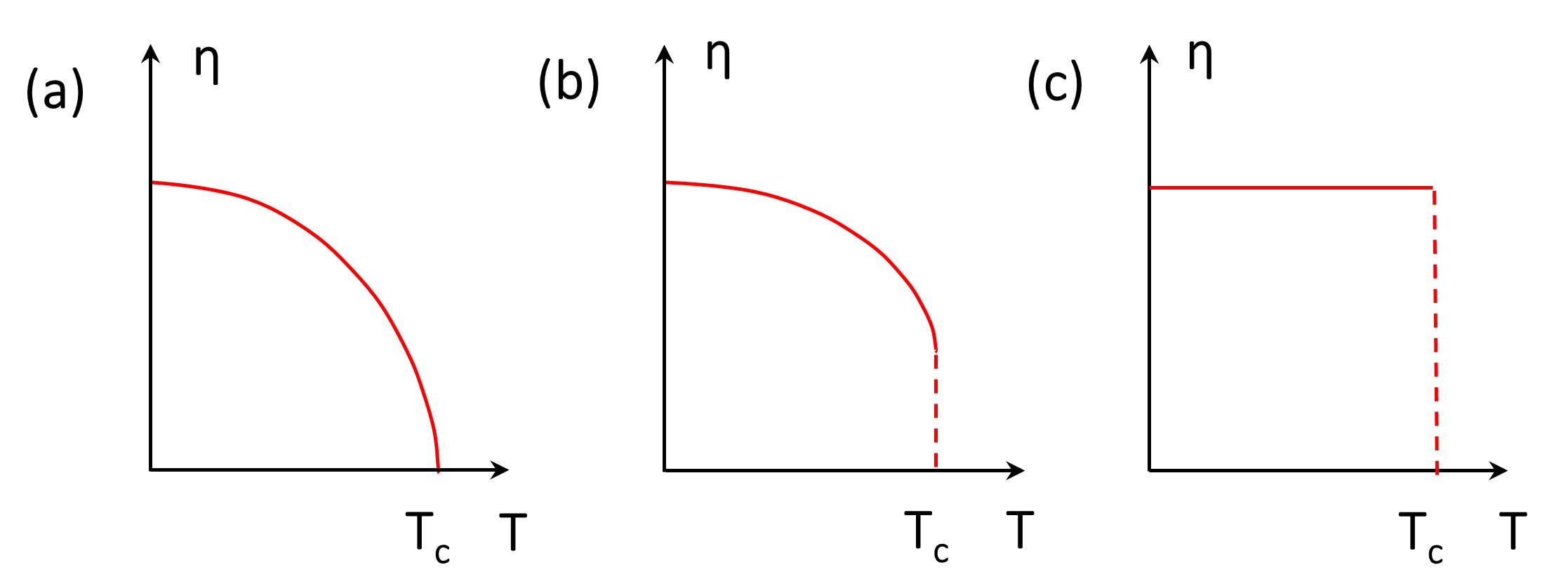}
\caption{(a) Temperature dependences of the order parameter for a second-order phase transition, (b) a first-order close to second-order phase transition, and (c) a first-order transition. This figure was reproduced from the book by Strukov et al. \cite{Strukov1998}.}
\label{figS8}
\end{figure}

The definition of the temperature dependence of the order parameter for the ferroelectric order transition is presented in Figure \ref{figS8}.

\subsection*{VI. Temperature dependence of EFG at the Fe site}

The corrected formula for \( V_{zz}(T) \) reported in our previous works (Eq. 2 \cite{Schell2022}, Eq. 8 \cite{Dang2022}) is:
\begin{equation}
V_{zz} = \xi \cdot P_S^2 = \xi \cdot \frac{\beta}{2\gamma} \left[1 + \sqrt{1 - \frac{4\chi_0^{-1} \gamma}{\beta^2} (T - T_0)}\right] \tag{s8}
\end{equation}

Or converted into the following formula using the relation \( T_C = T_0 + \frac{3}{16} \frac{\beta^2}{\chi_0^{-1} \gamma} \):
\begin{equation}
V_{zz} = \xi \cdot P_S^2 = \xi \cdot \frac{\beta}{2\gamma} \left[1 + \frac{1}{2} \sqrt{4 - 3 \frac{(T - T_0)}{(T_C - T_0)}}\right] \tag{s9}
\end{equation}

At \( T = T_C \):
\begin{equation}
V_{zz} (T_C) = \xi \cdot \frac{3\beta}{4\gamma}\tag{s10}
\end{equation}

At \( T = T_0 \):
\begin{equation}
V_{zz} (T = T_0) = \xi \cdot \frac{\beta}{\gamma}\tag{s11}
\end{equation}

\begin{figure}[h!]
\renewcommand{\thefigure}{S\arabic{figure}}
\centering
\includegraphics[width=12cm]{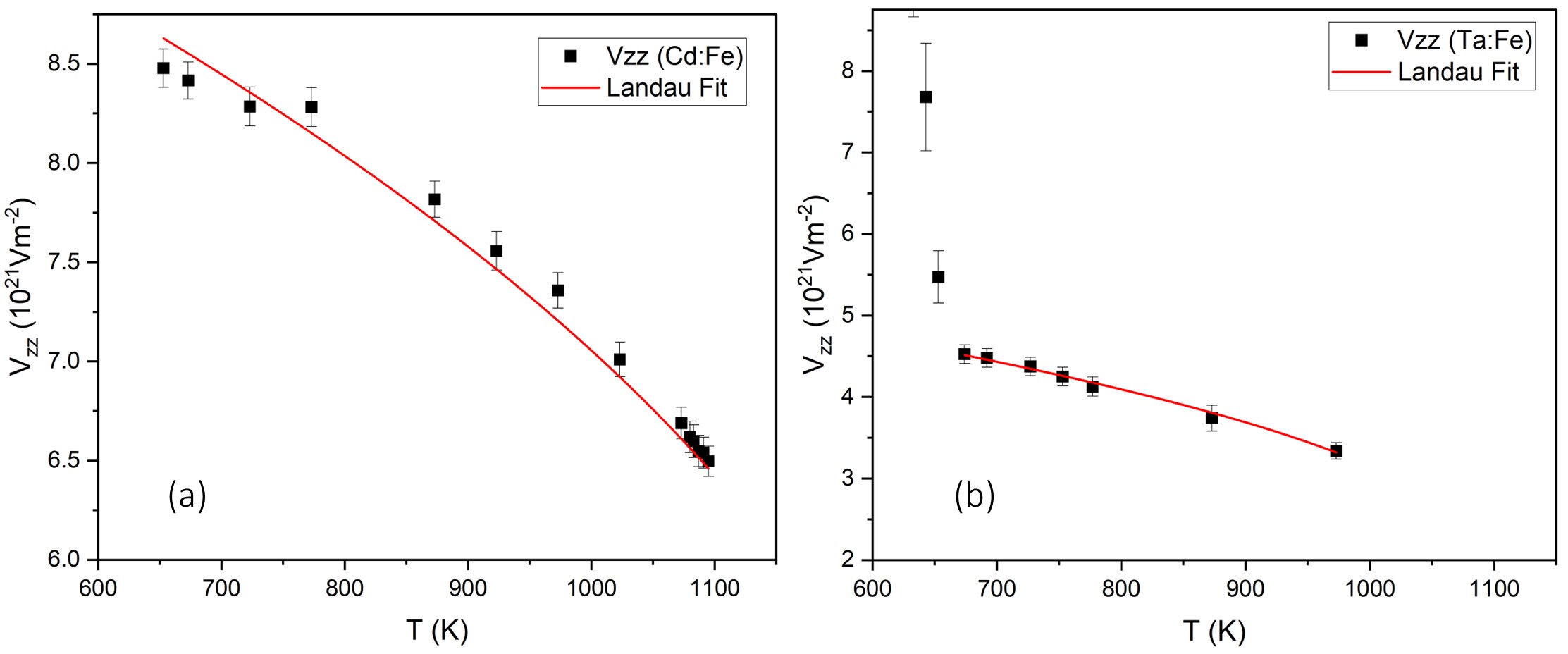}
\caption{(a) Temperature dependence of \( V_{zz} \) at the Fe site using \( ^{111}\)In \cite{Dang2022} and (b) \( ^{181}\)Hf probes \cite{Schell2022}. The data was re-fitted using the formula (s9) , which follows the Landau theory for the first-order phase close to the second-order phase transition \cite{Strukov1998, Lines1977}.}
\label{figS9}
\end{figure}

The formula (s8) was derived from Landau theory for a first-order close to second-order phase transition \cite{Strukov1998, Lines1977}. The formula (s9) was utilized to re-fit the data above the Néel temperature presented in Fig. 4 \cite{Schell2022} and Fig. 17 \cite{Dang2022}. The new fittings are presented in Fig. \ref{figS9}, and the extracted parameters are shown in Table III.

It was concluded in our previous publications \cite{Schell2022, Dang2022} that there is a huge coupling between the ferroelectric order and antiferromagnetic order of BiFeO$_3$ below the Néel temperature. This statement is corrected as follows: “there is a huge coupling between the antiferrodistortive (AFD) order and antiferromagnetic (AFM) order of BiFeO$_3$. The ferromagnetic (FM) order and AFM order are decoupled in bulk BFO.” See the main manuscript for this argument.

\begin{table}[h]
\renewcommand{\thetable}{S\arabic{table}}
\centering
\caption{The extracted parameters of the Landau fits for temperature dependence \( V_{zz} \).}
\begin{tabular}{ccccc}
    \hline
    & \(\xi\) [V C\(^{-2}\) m\(^{2}\)] & \(\beta/\gamma\) [C\(^{2}\) m\(^{-4}\)] & \(T_{0}\) [K] & \(T_{C}\) [K] \\
    \hline
    Ta:Fe & \(2.18 \times 10^{21}\) & 1.82 & 836(4) & 1026(7) \\
    Cd:Fe & \(2.84 \times 10^{21}\) & 2.76 & 843(8) & 1168(7) \\
    \hline
\end{tabular}
\end{table}

\end{document}